\documentclass[fleqn,usenatbib]{mnras}
\usepackage{graphicx} 
\usepackage{amsmath} 
\usepackage{amssymb} 
\usepackage{url}
\usepackage{ragged2e}
\usepackage{wrapfig}
\usepackage{textcomp}
\usepackage{hyperref}
\usepackage{setspace}
\usepackage{fancyhdr}
\usepackage{pdfpages}
\usepackage{float}
\usepackage{epstopdf}
\usepackage[caption=false]{subfig}
\usepackage[toc, page]{appendix}

\usepackage[caption=false]{subfig}

\title[CMB  Signal and Parameter Estimation]{A Foreground Model Independent Bayesian CMB Temperature and Polarization Signal Reconstruction and Cosmological Parameter Estimation over Large Angular Scales }

\author[Albin Joseph et al.]{Albin Joseph$^{1}$\thanks{Email: albinje@iiserb.ac.in},Ujjal Purkayastha$^{1}$\thanks{Email: ujjalp@iiserb.ac.in}, 
	Rajib Saha$^{1}$\thanks{Email: rajib@iiserb.ac.in}
	\\
	$^{1}$ Department of Physics, Indian Institute of Science Education and Research Bhopal -462066}

\date{Accepted XXX. Received YYY; in original form ZZZ}
\pubyear{2023}

\begin{document}
	\label{firstpage}
\pagerange{\pageref{firstpage}--\pageref{lastpage}}	
	\maketitle

\begin{abstract}

Recent CMB observations have resulted in very precise observational data. A robust and reliable CMB reconstruction technique can lead to efficient estimation of the cosmological parameters. We demonstrate the performance of our methodology using simulated temperature and polarization observations using cosmic variance limited future generation PRISM satellite mission. We generate samples from the joint distribution by implementing the CMB inverse covariance weighted internal-linear-combination (ILC) with the Gibbs sampling technique. We use the Python Sky Model ({\tt PySM)}, {\tt d4f1s1} to generate the realistic foreground templates. The synchrotron emission is parametrized by a spatially varying spectral index, whereas the thermal dust emission is described as a two-component dust model. We estimate the marginalized densities of CMB signal and theoretical angular power spectrum utilizing the samples from the entire posterior distribution. The best-fit cleaned CMB map and the 
corresponding angular power spectrum are consistent with the CMB realization and the sky angular power spectrum, implying an efficient foreground minimized reconstruction.
The likelihood function estimated by making use of the Blackwell-Rao estimator is used for the estimation of the cosmological parameters. Our methodology can estimate the tensor to scalar ratio $r\ge 0.0075$ for the chosen foreground models and the instrumental noise levels.
Our current work demonstrates an analysis
pipeline starting from the reliable estimation of CMB signal and its angular power spectrum to the case of cosmological parameter estimation using the foreground model independent Gibbs-ILC method.

\end{abstract}


\maketitle

\section{Introduction}

The cornerstone of modern observational cosmology has been the accurate measurement of the CMB signal and, thereby, the extraction of cosmological information. Over the last two decades, CMB anisotropies have been studied with excruciating detail. It has established $\Lambda$CDM  concordance model on a firm footing \citep{2020A&A...641A...6P}. One of the crucial benefits of analyzing CMB signal is that it can be used to constrain fundamental cosmological parameters \citep{2020A&A...641A...6P,2022MNRAS.tmp.3359J,2022MNRAS.511.1637J}. Temperature and $E$-mode fluctuations originate due to the primordial quantum fluctuations. Polarized CMB $E$-mode has been a useful tool to probe the reionization epoch \citep{PhysRevD.55.1830}. Accurate measurements of the CMB $E$-mode signal can break the degeneracy between the amplitude of the primordial power spectrum and the optical depth to reionization~\citep{2016ASSL..423.....M}. Constraining the reionization optical depth parameter $\tau$ can unravel the physics of early reionization~\citep{2003CRPhy...4..917K}. CMB polarization  can be a useful tool to probe the early star formation~\citep{10.1093/mnras/stx306,Michael_Shull_2012}. Although weak,  $B$-mode polarization anisotropies of CMB can serve to constrain various physical processes in the primordial era when the energy was around $10^{16}$GeV. The CMB $B$-mode polarization can also serve as a direct test of slow-roll inflation ~\citep{2022arXiv220316556B}. The signature of the inflationary gravitational wave can be measured by CMB $B$-mode, which in turn can also be used to constrain the energy scale of inflation ~\citep{PhysRevLett.78.1861}. 
The above discussions show that to understand cosmological physics, one has to analyze the CMB signal devoid of any impurities. To reliably estimate the cosmological parameters from the CMB  signal on large angular scales, we require foreground mitigation techniques that can provide accurate estimates of the signal along with the statistical uncertainties.

Statistical analysis of the CMB can be carried out by the joint posterior distribution $P({\bf S}, C^X_\ell |{\bf D})$ of the CMB  signal ${\bf S}$ and fiducial angular power spectrum $C^X_\ell$ given the observed CMB maps. Here $X$ can be any CMB field $T,E, B$ and  ${\bf D}$ is the observed data. Following the Bayesian argument, one can conclude that all the information is contained in the posterior distribution. This posterior density can be employed to find the best-fit ${\bf S}$ and $C^X_\ell$ along with their associated error bars. One can then estimate the likelihood function $P(C^X_\ell | {\bf D})$ by marginalizing the posterior over the CMB signal which plays a central role in estimating the cosmological parameters. The foreground cleaning methodology and signal reconstruction have a direct bearing on the estimation of the joint posterior density and the likelihood function since different foreground mitigation techniques will result in different posterior density evaluations. 
The precise estimation of likelihood functions is crucial for the CMB reconstruction and the correct interpretation of cosmological parameters. In this article, we estimate the joint posterior density, and the  likelihood function of the cleaned CMB maps over large angular scales using our inverse covariance weighted internal linear combination (ILC) incorporating the Gibbs sampling technique \citep{10.1214/ss/1177011136,PhysRevD.71.103002,doi:10.1142/S0217732322500997}. Moreover, to efficiently estimate cosmological parameters, we employ the likelihood function using all the $X$ fields of $C^X_\ell$.

In order to determine the joint conditional density and the likelihood function, as a first step, one has to generate the foreground minimized CMB maps. The major foreground component at frequencies $\lesssim 70$ GHz ~\citep{2020A&A...641A...4P} is the synchrotron radiation resulting from the accelerated motion of cosmic ray particles in the galactic magnetic field. The thermal dust foreground emission plays a dominant role above $100$ GHz\citep{2020A&A...641A...4P}. Free-free, also known as bremsstrahlung, originates from the electron-ion collision at interstellar plasma and is a major source of foreground at frequencies $10 - 100$ GHz~\citep{10.1046/j.1365-8711.2003.06439.x}. Free-free is intrinsically unpolarized and hence does not pose as a  polarized foreground. Apart from these foregrounds, reconstruction can also be hampered by the presence of detector noise. Reconstruction of weak polarized  CMB signal by mitigating the strong foregrounds in presence of noise therefore becomes a challenging task. Nevertheless, many future CMB missions are being designed to accurately measure the CMB $B$-mode fluctuations with a sufficiently large signal-to-noise ratio. The $B$-mode signal is most susceptible to residual noise bias owing to its weak nature. Therefore, we perform the noise bias correction before estimating the cleaned $B$-mode angular power spectrum.    

Foreground removal and CMB reconstruction can be performed using the following two methodologies. The first approach minimizes the contribution from all astrophysical components while preserving the CMB signal without using any explicit information about the foreground models. Here the only assumption is that foregrounds do not follow the black-body nature, whereas the CMB does. The second approach requires the knowledge of model parameters, frequency dependence and (or) spectral energy distribution of the different foreground components present in the microwave sky. These techniques are referred to foreground model-dependent methodology viz. Wiener filtering~\citep{1999MNRAS.302..663B}, Gibbs sampling approach~\citep{2009ApJ...690.1807G,2008ApJ...676...10E,2008ApJ...672L..87E,2004PhRvD..70h3511W}, template fitting method~\citep{2012MNRAS.420.2162F}, the maximum entropy method~\citep{Gold_2009}, and  { Markov Chain Monte Carlo}~\citep{Gold_2009} method. The above mentioned algorithms work based on external information where one exploits the spectral modeling of all the sky components. On the other hand, the former requires a minimal assumption about the foreground components, and one can solely focus on CMB reconstruction and analysis. Various model-independent methods have been mentioned in the literature such as Independent Component Analysis (ICA)~\citep{2003MNRAS.344..544M,2002MNRAS.334...53M,2010MNRAS.402..207B}, ILC~\citep{10.1093/mnras/281.4.1297}, and Correlated Component Analysis (CCA)~\citep{2005EJASP2005.2400B}.

ILC is a foreground reduction algorithm that relies on a simple yet reliable assumption that the foreground and noise are non-blackbody in nature in contrast to CMB. The cleaned CMB map produced by the ILC approach is ``robust" to foreground modelling inaccuracies. In the ILC method, a foreground reduced  CMB map is obtained by combining weighted multi-frequency observed foreground contaminated CMB maps. These weights are subjected to the signal preserving constraint that the weights of all frequency bands sum to unity. The dependency of foreground minimization and cross-correlation effects over the number of frequency channels and the total number of components is reported in~\cite{2009MNRAS.397.1355E} and ~\cite{PhysRevD.78.023003}. One can also do an ILC-type likelihood estimation using the CMB covariance~\citep{2008arXiv0805.0093G}. In the  ILC method, these weights are computed by Lagrange's method of minimization of the variance of the cleaned CMB map. The analytical nature of the estimation of the weights is an added advantage of the method since the numerical minimization algorithm may suffer from convergence issues.

Since we use the ILC method for  CMB reconstruction, in the current article, the posterior density and 
the likelihood function estimated are free from inaccurate modelling of the foregrounds. Thus, our methodology is an important and complementary addition to the existing foreground model-dependent cleaning of CMB.   Using a large number of input frequency maps in the analysis results in negligible residual foregrounds ~\citep{PhysRevD.78.023003} and hence 
efficient reconstruction of the cleaned map and angular power spectrum. For a more detailed analysis of the computation of joint posterior distribution using explicit models of foreground components, one can refer to~\cite{Eriksen_2008}.

The ILC methodology has been studied rigorously in~\cite{10.1111/j.1365-2966.2011.19770.x},~\cite{PhysRevD.68.123523},~\cite{SOURADEEP2006854},~\cite{Eriksen_2008},~\cite{2008A&A...491..597L},~\cite{PhysRevD.77.103002},~\cite{Samal_2010},    and~\cite{2009A&A...493..835D}. A global ILC technique was suggested~\citep{2018ApJ...867...74S} where the weights are determined by minimizing a CMB inverse covariance weighted variance rather than the typical variance in the cleaned maps. In~\cite{10.1093/mnras/staa3935,doi:10.1142/S0217732322500997}, this methodology has been extended to estimate a foreground reduced CMB $E$-mode map at large angular scales. \textcolor{black}{In the current work, we demonstrate an analysis
	pipeline starting from the reliable estimation of CMB signal and its angular power spectrum to the case of cosmological parameter estimation using the foreground model independent Gibbs-ILC method. It is desirable in the CMB component analysis we use a CMB signal reconstruction which can provide the error estimates on the estimated signal and angular power spectrum. In our current approach, we use a novel method, namely the Gibbs-ILC method, which has twofold advantages. First of all, the methodology is foreground model-independent, and in this framework, we can also estimate the joint posterior distributions of the cleaned CMB map and its angular power spectrum. Since the Gibbs-ILC method provides the probability density functions of signal and angular power spectrum, in this current work, we utilize them for cosmological parameter estimation. This is a unique foreground cleaning approach along with some Bayesian approaches like Gibbs-ILC, which will provide the estimates of error on the reconstructed signal and angular power spectrum, which can be used for cosmological analysis. An added advantage is that the induced errors are nicely propagated in the final cosmological parameter estimation. As shown in~\cite{PhysRevD.78.023003}, if we have sufficient number of frequency channels and provided the detector noise is negligible, we are in a position to reliably remove the foregrounds accurately. We would also like to mention that the foreground removal and parameter estimation are modular in nature so that, if required, they can be modified as need be raised.}

We have organized the paper as follows. In section~\ref{formalism}, we review the basic formalism of our
method. In section~\ref{input maps}, we discuss the input maps that we have used in our analysis and the methodology in section~\ref{Methods}. We then describe the results of
our analysis in section~\ref{Sec:Results}. After estimating the parameters from all three cleaned power spectra in section~\ref{br}, we discuss and draw our conclusions in the last section~\ref{disc}.

\section{Formalism} \label{formalism}

CMB temperature anisotropy $\Delta T$ at a given direction $\hat{n}$  can be represented as a scalar field.  One can  therefore decompose the temperature fluctuations  on a surface of a sphere  by spin-0 spherical harmonics as,
\begin{equation}
	\frac{\Delta T(\hat{n})}{T} = \sum_{\ell=2}^{\ell_{max}}a_{\ell m} Y_{\ell m} (\hat{n} ) ,
\end{equation}
where $T$ is the isotropic CMB temperature. Thomson scattering induces linear polarization in the CMB temperature fluctuations. These can be conveniently represented by Stoke's $Q$ and $U$ parameters. However, the linear combinations $Q\pm iU$ transform as spin-2 objects given by,
\begin{equation}
	Q(\hat{n})\pm iU(\hat{n}) = \sum_{\ell=2}^{\ell_{max}}a_{\pm 2, \ell m} Y_{\pm2,\ell m} (\hat{n} ) ,
\end{equation} where $ Y_{\pm2,\ell m} (\hat{n})$ denotes the $\pm 2$ spherical harmonics. One can construct spin-0 $E(\hat{n})$  and $B(\hat{n})$  map by suitable linear combinations of spin-2 spherical harmonic coefficients as
\begin{equation}
	E(\hat{n}) = \sum_{\ell=2}^{\ell_{max}}a^{E}_{\ell m} Y_{\ell m} (\hat{n} ),
\end{equation}
\begin{equation}
	B(\hat{n}) = \sum_{\ell=2}^{\ell_{max}}a^{B}_{\ell m} Y_{\ell m} (\hat{n} ) ,
\end{equation}
where $a^{E}_{\ell m} = (a_{2,\ell m}+a_{-2,\ell m})/2$ and $a^{B}_{\ell m} = (a_{2,\ell m}-a_{-2,\ell m})/2i$. Since the conversion $ (Q,U)$ to $ (E,B)$ is over the entire sky, one can always reconstruct $ (Q,U)$ from $ (E,B)$. Furthermore, a full sky conversion also prevents the problem of leakage from $E$ to $B$.

Let us consider the sky signal observed over a full microwave sky. The observed map at a frequency $\nu$ is then given by 
\begin{equation}
	\mathbf{d_{\nu}} = \mathbf{S} + \mathbf{F_{\nu}} + \mathbf{N_{\nu}}.
	\label{sky}
\end{equation}
Here $\mathbf{S}, \mathbf{F_{\nu}}$ \text{and} $\mathbf{N_{\nu}}$ denotes signal, foreground, and the detector noise contribution for the $\nu^{th}$ frequency, respectively. The bold-faced quantities in Eqn.~\ref{sky} denote column vector of size ${N}_{\text{pix}} \times 1 $ column vector. Here ${N}_{\text{pix}} = 12{N}_{\text{side}}^2$ denotes the number of pixels in a map and ${N}_{\text{side}}$ represents the HEALPix  pixel resolution parameter  vector. The observed data ${\bf D} = \{{\bf d}_1, {\bf d}_2,...,{\bf d}_n \})$
can be represented by a matrix of size $N_{\text{pix}} \times n$. Since CMB follows the black body spectrum, $\mathbf{S}$ is independent of frequency $\nu$. The noise contamination for the PRISM satellite mission can be considered negligible in comparison to the temperature and $E$-mode signal. However, for the weak $B$-mode signal, residual noise can bias the signal reconstruction. For a more detailed discussion about noise bias minimization, we refer to Eqn.\ref{bias}.

\begin{figure}
	\centering
	\subfloat{\includegraphics[width=.8\linewidth]{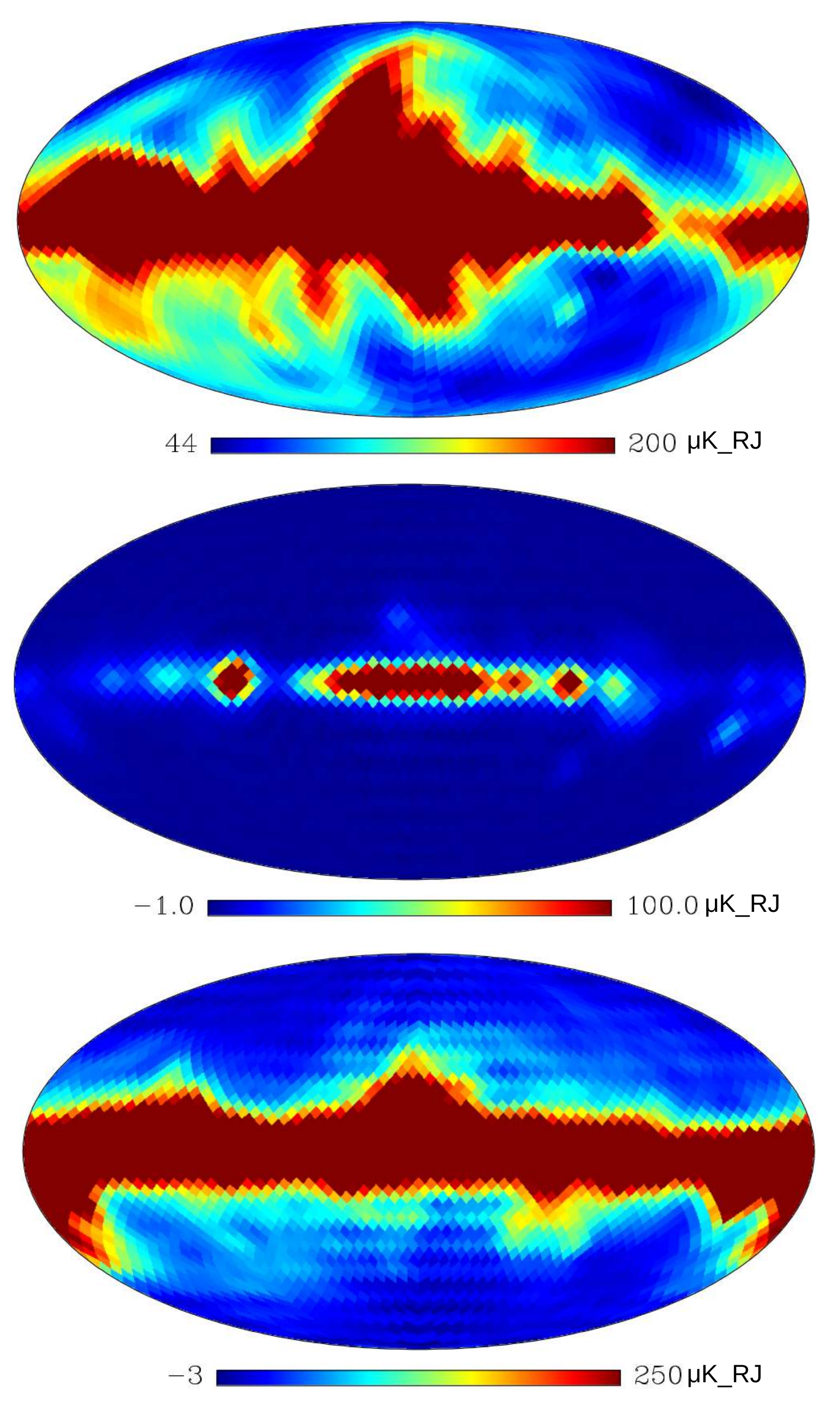}}
	\caption { PySM {\tt d4f1s1} temperature foreground  templates in the units of $\mu$K\textunderscore
		RJ. The top panel shows the synchrotron emission map at 21 GHz and the free-free template at 155 GHz is illustrated in the middle panel. The bottom panel represents the thermal dust template at 385 GHz. From the figure, it is evident that the galactic plane exhibits strong foreground emission.}\label{Tfg}
\end{figure}

\begin{figure}
	\centering
	\subfloat{\includegraphics[width=.7\linewidth]{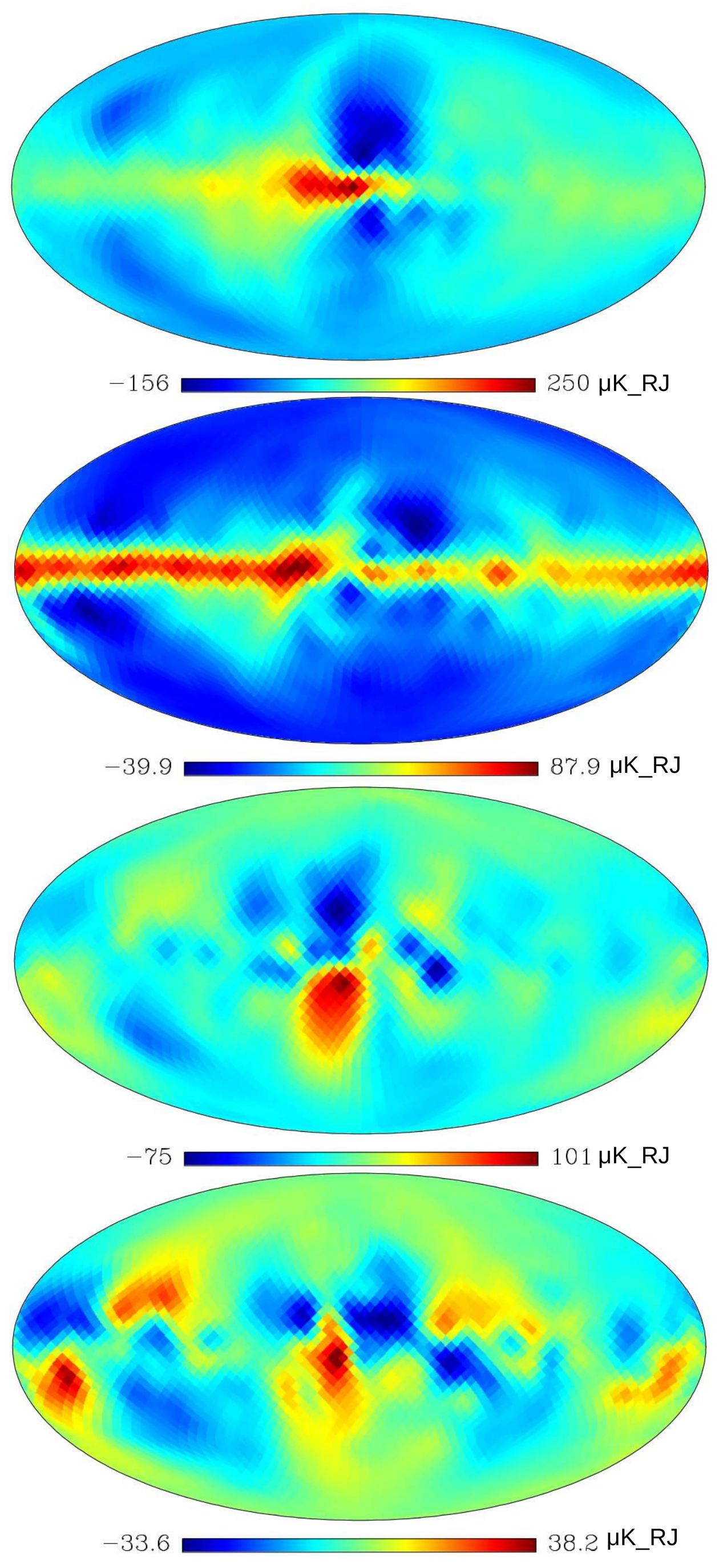}}
	\caption { PySM {\tt d4s1} foreground polarization templates in the units of $\mu$K\textunderscore
		RJ. The first and second panel shows  $E$-mode synchrotron at 21 GHz and $E$-mode thermal dust template at 799 GHz, respectively. The third and fourth panels are the same as the first two panels but for $B$-modes. The $E$-mode foreground intensity is strong mainly across the galactic plane, whereas for the $B$-mode, the morphological pattern is different, and the foregrounds dominate the upper and lower halves of the galactic plane.}\label{EBfg}
\end{figure}


To estimate the posterior density  $P({\bf S}, C^X_\ell \vert {\bf D})$ of the CMB field, where  $X=T,E,B$, one can draw a large number of samples by a direct evaluation of a known posterior distribution. Alternatively, a powerful technique that can be utilized is so-called Gibbs sampling, which works on the principle that the samples are drawn from the two conditional distributions that are straightforward to sample. The Gibbs $(k+1)^{th}$  CMB  signal for any of the field $X$ is obtained  by drawing sample from \begin{equation}
	{\bf S}^{k+1} \leftarrow P_1({\bf S}|{\bf D},C_{\ell}^{X,k}) \, ,
	\label{cmbsamp}
\end{equation} 
where $C_\ell^{X,k}$ denotes the corresponding theoretical angular power spectrum computed at the previous 
step. Sampling the conditional density $C_\ell^{X,k+1}$ yields  
\begin{equation}
	C_\ell^{X,k+1} \leftarrow P_2(C^{X,k}_{\ell}|{\bf D},{\bf S}^{k+1}) \, .
	\label{clsamp}
\end{equation}
Eqn.~\ref{cmbsamp} and  Eqn.~\ref{clsamp} are repeated until the chain converges. The samples from the joint posterior densities are estimated from the samples obtained from the two conditional distributions after initial burn-in rejection. A natural question arises:  how does one sample CMB signal given the data $\mathbf{D}$  and theoretical angular power spectrum $C^X_{\ell}$? This is accomplished by applying the foreground removal approach proposed by~\cite{Sudevan_2022},~\cite{10.1093/mnras/staa3935}, and~\cite{doi:10.1142/S0217732322500997} to estimate the cleaned  CMB signal given the data and a theoretical CMB angular power spectrum. A foreground reduced cleaned map can be obtained by a linear combination of all the available $n$ input maps in the pixel domain following the usual ILC algorithm,
\begin{equation}
	{\bf  S} = \sum_{i=1}^{n} w_i {\bf d}_i = \sum_{i=1}^{n} w_i \sum_{\ell > 0}\sum_{m=-\ell}^{\ell} a_{\ell m}^i Y_{\ell m}({\bf \hat{n}})\, ,
	\label{weight}
\end{equation}
where $w_i$ represents the weight corresponding to the map from $i^{th}$ frequency channel. In order to ensure that the CMB signal will not undergo any undesired modification during the entire foreground-removal procedure, the weights are subject to the constraint that they sum to unity, i.e., $\sum_{i=1}^{n} w_i = 1$.
	Furthermore, the conditional density $P({\bf S}|{\bf D},C_{\ell}^{X,i})$ can be obtained by maximizing the likelihood of the model given the full-sky observed CMB data and the theoretical angular power spectrum in the spherical harmonic domain as follows~\citep{Sudevan_2022}:
\begin{eqnarray}
	P_1({\bf S} \vert {\bf D}, C_{\ell}^X)  \propto \prod_{\ell, m} e^{- \sum\limits_{i,j} w_{i}a_{\ell m}^{i}a_{\ell m}^{\ast j}w_j / C_{\ell}^X}  \\ \propto   e^{-\sum\limits_{i,j}w_{i}w_j {\sum\limits_{\ell} (2\ell + 1) \hat{\sigma}_{\ell}^{ij} / C_{\ell}^X} } \, .
	\label{hilc_new}
\end{eqnarray} 
Here the  ${\hat{\sigma}_\ell^{ij}}$ is the cross-power spectrum between the observed CMB maps ${\bf d}_i$ and ${\bf d}_j$. Using Eqn.~\ref{hilc_new}, we can define an estimator $\sigma^2$ given by,
\begin{equation}
	\sigma^2  = \sum_{i}^{}\sum_{j}^{}w_{i}w_{j}\sum_{\ell=2}^{\ell_{\textnormal{max}}}(2\ell+1)\frac{\hat{\sigma}^{ij}_\ell}{C^{X^\prime}_{\ell}}  \, .
	\label{aij1}
\end{equation}
We then minimize the $\sigma^2$ and obtain the weights to sample the CMB signal. In the previous equation, $C_\ell^{X^\prime}$ denotes  the appropriate  beam and pixel convolved CMB  theoretical power spectrum such that,
\begin{equation}
	{C_{\ell}^{X^\prime}} = {C^X_{\ell}{B^2_{\ell}} {P^2_{\ell}} } \, ,
\end{equation}
where $C^X_\ell$  is free of any smoothing effects. We employ Lagrange's multiplier approach, to minimize $\sigma^2$ and solve for the weights as,
\begin{equation}
	{\bf W} = \frac{{\bf {\hat A}^\dagger} {\bf e}}{{\bf e}^T {\bf {\hat A}^\dagger} {\bf e}}\, .
	\label{weightnew}
\end{equation}
Here {\bf W} is a $(n \times 1)$ weight vector and {\bf e} is a $(n \times 1)$ shape vector of CMB in thermodynamic units. The ($i, j$) elements of the matrix $\bf A$ can be represented in the pixel space as,
\begin{equation}
	\hat{A}_{ij} = {\bf d}_i^T {\bf C}^\dagger {\bf d}_j\, ,
	\label{mat}
\end{equation}
where $\bf C$ denotes CMB theoretical covariance matrix and $\dagger$ represents Moore-Penrose generalized inverse~\citep{Sudevan_2018, penrose_1955}. Since the computation in pixel space is intensive, one can conveniently evaluate Eqn.~\ref{mat} in the harmonic domain as~\citep{Sudevan_2020},
\begin{equation}
	{\hat A}_{ij} = \sum_{\ell=2}^{\ell_{\tt max}} (2\ell + 1) \frac{\hat{\sigma}_\ell^{ij}}{C_\ell^{X^\prime}} \, . 
	\label{hilc1}
\end{equation}
We then use the matrix $\bf A$, the elements of which are given in Eqn.~\ref{hilc1}, in Eqn.~\ref{weightnew} to obtain the weight row vector $w$. Furthermore,  we use the obtained weights in Eqn.~\ref{weight}, to sample the foreground minimized CMB signal ($T, E, B$),  $\bf S$ by linearly combining the input channel maps $\bf d_i$.

Once we obtain the sample of CMB signal $\bf S$ given the data $\bf D$ and $C_{\ell}^X$, we proceed to the second step of Gibbs sampling where we draw samples of $C_{\ell}^X$ given $\bf S$ and $\bf D$. The signal sample $\bf S$, obtained from the initial step of Gibbs sampling can be represented as,
\begin{equation}
	\textbf{S}(\theta,\phi) = \sum_{\ell=2}^{\infty}\sum_{m=-\ell}^{\ell}s_{\ell m}\textbf{Y}_{\ell m}(\theta,\phi) .
	\label{eq12}
\end{equation}
So the realization specific power spectrum can be  denoted by,
\begin{equation}
	\hat{\sigma}_{\ell}^{X^{}} = \frac{1}{2\ell +1}\sum_{m=-\ell}^{\ell} s^2_{\ell m}\,,
	\label{specific}
\end{equation}
where $X= T, E, B$. Since the $B$-mode signal is the weakest of the three types of signals analyzed in this work, some specific noise debiasing is required from the cleaned $B$-mode power spectrum before using it to sample an estimate of the theoretical CMB $B$-mode power spectrum. To perform this, we first subtract the noise bias of each cleaned $B$-mode map by computing its noise level~\citep{Yadav_2021}. Thus the debiased cleaned CMB $B$-mode power spectrum can be obtained by,
\begin{equation}\label{bias}
	\hat{\sigma}_{\ell}^{B'^{}}= \hat{\sigma}_{\ell}^{B^{}}  -  \sum_{i=1}^{n} w^2_{i}\sigma_{\ell}^{N,i}\,.
\end{equation}
Here, the second term on the right-hand side represents the noise levels on each cleaned CMB $B$-mode map, with $\sigma_{\ell}^{N,i}$ and $w_i$ denoting the noise auto power spectrum and weights corresponding to $i^{th}$ frequency channel, respectively. After the noise bias subtraction our debiased theoretical $B$-mode power spectrum is given by $\hat{\sigma}_{\ell}^{B'^{}}$. However, for the purpose of common usage of notations, we represent debiased CMB $B$-mode power spectrum by 	$\hat{\sigma}_{\ell}^{B^{}}$.


\begin{figure*}
	\centering
	\includegraphics[scale=0.66,angle =0]{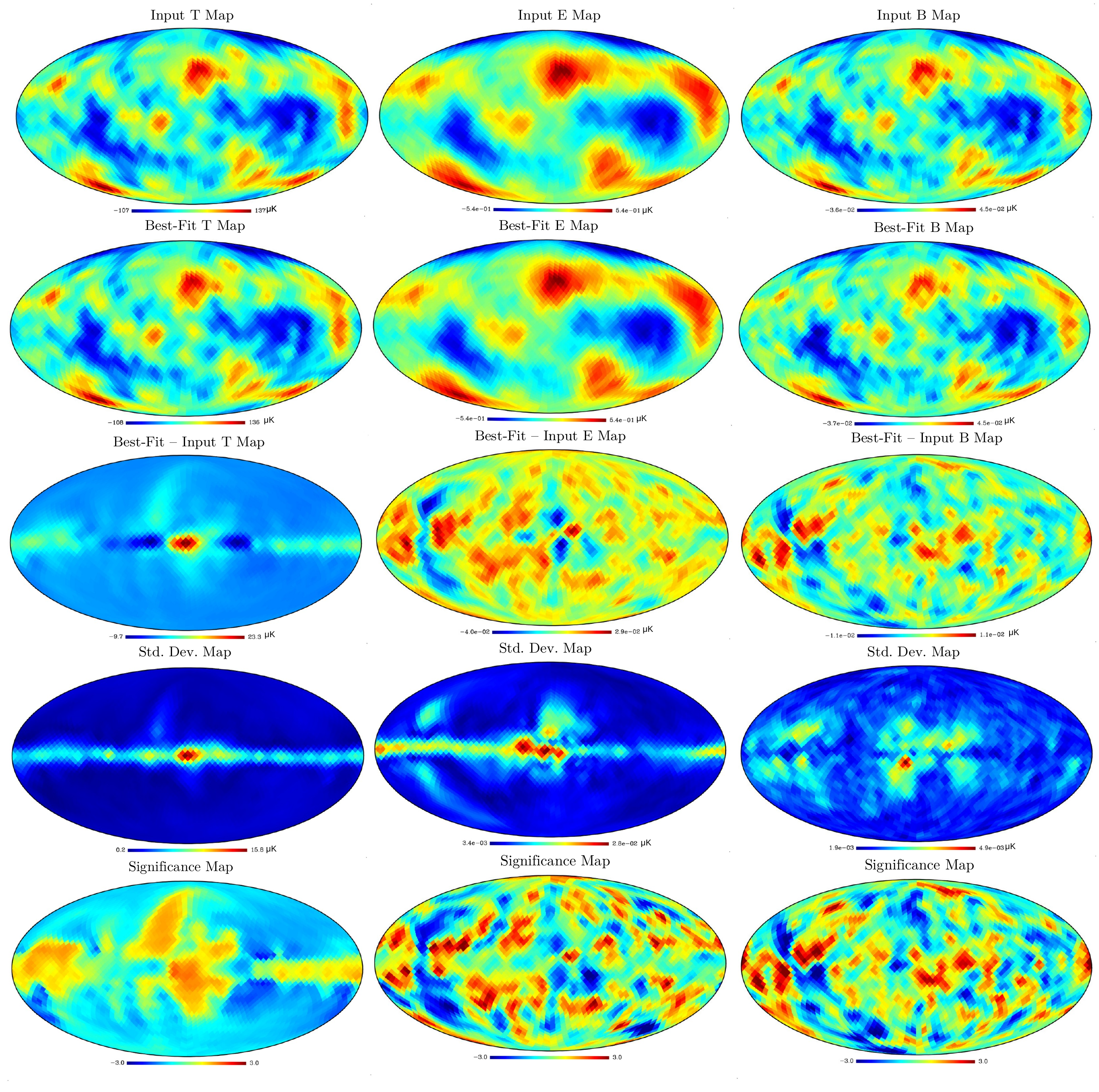}
	\caption{In the top panel, we show the input CMB T, E and B maps respectively, for an arbitrarily chosen realization. The best-fit cleaned  CMB T, E and B maps obtained from the marginalized value of the histogram for the same input realization are shown in the second panel. Both the input maps and the best-fit maps of CMB T, E and B-mode agree well with each other. The difference maps (residuals) obtained by subtracting the input maps from the corresponding best-fit cleaned maps are shown in the middle panel. In the fourth panel, we show the standard deviation maps obtained from all the difference maps of the 200 simulations. The standard deviation map for the weakest B-mode has residual only of the order $\lesssim 0.0049 \mu K$. Apart from some of the pixels in the central region and towards the left (for E-mode and B-mode) we can conclude that an efficient foreground removal has been achieved. The last panel shows the standard deviation weighted residuals in units of $\pm 3\sigma$. }
	\label{cleanedmapsB}
\end{figure*}
Alternatively, the cross power spectrum may be utilized to generate noise bias free power spectrum~\citep{2007ApJS..170..288H}. The limitation of this method is that it would increase the noise variance by a factor of 2 in the input map. We, therefore, do not pursue this method in this work.
	In order to draw samples of $C^X_\ell$ given ${\bf S}$ and ${\bf D}$ we first 
	obtain the conditional density  $P_2(C^X_\ell|\textbf{S},\textbf{D})$
	in terms of the variable $y=\hat{\sigma}^X_\ell (2\ell+1)/{C^X_\ell}$ as \citep{Sudevan_2020},
\begin{eqnarray}
	P_{2}(y|\hat{\sigma}^X_\ell) \propto y^{-(2\ell-1)/2-1}  \text{exp}\left[-\frac{y}{2} \right] \, ,
	\label{eq_y}
\end{eqnarray}
where $\hat{\sigma}^X_\ell$ is estimated from the cleaned CMB  map as shown in Eqn.~\ref{specific} (note: for $B$-mode, the $\hat{\sigma}^B_\ell$ is obtained after an additional bias correction as given in Eqn.~\ref{bias}). Furthermore, from Eqn.~\ref{eq_y} it is evident that the variable  $y$ follows a $\chi ^2$ distribution with $ 2\ell-1$ degrees of freedom (dof). Hence,  to sample a CMB $C_{\ell}^X$, we draw
	$y$ from the $\chi^2$ distribution of $2\ell-1$ dof and subsequently we compute $C^X_\ell$ as,
\begin{equation}
	C^X_\ell = \hat{\sigma}^X_\ell (2\ell + 1)/y.
\end{equation}
However, we note that the distribution of $C_{\ell}^X$ is an inverse-gamma distribution given by,
\begin{eqnarray}
	P_{2}(C^X_{\ell}|\hat{\sigma}^X_\ell) \propto \left(\frac{1}{C_{\ell}^X}\right)^{\left(2\ell+1\right)/2}
	\exp\left[-\frac{\hat{\sigma}_{\ell}^X\left(2\ell+1\right)}{2C_{\ell}^X}\right]\,.
	\label{pdf_cl}
\end{eqnarray}

\section{Input Maps}
\label{input maps}
In this work, we generate fixed foreground templates and randomly simulated noise maps at the Polarized Radiation Imaging and Spectroscopy Mission (PRISM)  frequency channels ranging from $21$ GHz to $799$ GHz. These maps are then added to the respective CMB simulations. The input frequency bands used in our work along with their respective beam and noise sensitivities are shown in Table~\ref{tab:prism}.

\subsection{CMB Maps}

We generate random Gaussian realizations of the Stokes CMB $T$,  $Q$ and $U$   maps from lensed CMB $TT, EE$ and $BB$ angular power spectrum respectively using the Boltzmann solver CLASS~\citep{2011JCAP...07..034B,2011arXiv1104.2932L}; at HEALPix (Hierarchical Equal Area IsoLatitude Pixellation of sky)~\citep{Gorski_2005} resolution $N_{\text{side}} =16$. The $Q$ and $U$ maps are then converted to the CMB $E$-mode and $B$-mode maps employing the {\tt synfast} facility, and all the resulting maps are convolved by a Gaussian beam window of FWHM $9^\circ$. Since we perform the conversion over the full sky, the resulting $E,B$ maps do not suffer from leakage. To generate the CMB $TT$ and $EE$ power spectrum, we implement the cosmological parameter values from the Planck 2018 results~\citep{2020A&A...641A...6P} in the latest
version of the cosmological Boltzmann integrator code CLASS~\citep{2011JCAP...07..034B,2011arXiv1104.2932L}. To compute the $B$-mode power spectrum, in addition to the above-mentioned parameters we fix the tensor to scalar ratio $r=7.5\times10^{-3}$ and the lensing amplitude $A_{lens}=1.0$. From the latest {\tt NPIPE}-processed Public Release 4 (PR4) of temperature and polarization maps, the Planck collaboration~\citep{2021A&A...647A.128T} reported an upper limit of $r < 0.056$ at $95\%$ confidence level (C.L.). The tightest constraints to date on the $B$-mode with $r < 0.037$ at $95\%$ C.L. is obtained in~\cite{2022arXiv220505617C}, where they integrated Planck PR4 data with BICEP/Keck Array 2018, Planck CMB lensing and BAO using a frequentist profile likelihood method. Thus our choice of the tensor to scalar ratio $r=7.5\times10^{-3}$ is well below its current bound.

\subsection{Foreground model}
\label{fg}

Synchrotron, free-free and thermal dust are major astrophysical foreground emissions that corrupt the CMB temperature signal. Synchrotron and thermal dust emissions are known to be polarized and hence are the major limitations for the study of the CMB polarization signal. Free-free is intrinsically unpolarized. We use the publicly available software package Python Sky Model ({\tt PySM}) \citep{2017MNRAS.469.2821T} to generate the {\tt d4s1} templates of thermal dust and synchrotron for each of $T$ and polarized $Q$ and $U$ Stokes parameters. In addition to the above two components, for temperature, we generate the free-free template using the {\tt PySM} {\tt f1} model. A more detailed description of the individual foreground components is discussed below. In Fig.~\ref{Tfg} and Fig.~\ref{EBfg}, we show temperature as well as $E, B$ foreground templates for some of the PRISM frequency channels in the $\mu$K\textunderscore RJ unit.
\subsubsection{Synchrotron {\tt s1}}

To generate the synchroton emission template, 408-MHz Haslam map is used following \cite{2015MNRAS.451.4311R}.  The polarization modelling is obtained using the WMAP Stokes $Q$ and $U$ maps at 23 GHz \citep{2013ApJS..208...20B}.  The {\tt s1}  model of the {\tt PySM} assumes a power law behaviour with a spatially varying spectral index $\beta(\hat{n})$ at WMAP reference frequency $\nu_{0} = 23$ GHz given by,
\begin{equation}
	I^{synch}_{\nu}(\hat{n}) = A_{\nu_{0}}(\hat{n})\left( \frac{\nu}{\nu_{0}} \right)^{\beta(\hat{n})}\,.
\end{equation}
The spectral index map {\tt s1} in {\tt PySM} is taken from `Model 4' of \cite{2008A&A...490.1093M}.

\subsubsection{Free-free {\tt f1}}

The {\tt PySM} free-free emission template is unpolarized and obeys a power-law with a constant spectral index of -2.14 which flattens abruptly at lower frequencies. It uses degree-scale smoothed emission measure and effective electron temperature {\tt Commander} templates~\citep{2016A&A...594A..10P}. The spectral index chosen is in accordance with WMAP and Planck measurements for electrons $\sim 8000K$~\citep{2016A&A...594A..10P}.


\subsubsection{Thermal dust  {\tt d4}}

Thermal dust emission is modelled as two-component dust models~\citep{1999ApJ...524..867F}. Instead of using single-Modified Black Body (MBB) models, we considered the two-component dust models, which provide a better fit than the former~\citep{2015ApJ...798...88M}. The two-component dust models are parameterized by their own temperature and spectral indices template~\citep{2017MNRAS.469.2821T}:
\begin{equation}
	I^{dust}_{\nu}(\hat{n}) = \sum_{\alpha}^{2}I_{\alpha} (\hat{n})\left( \frac{\nu}{\nu_{0}} \right)^{\beta_{\alpha}} \frac{ B_{\nu}(T_{\alpha} (\hat{n}))}{B_{\nu*}(T_{\alpha} (\hat{n}))} ,
\end{equation}
where $\nu* = 545$ GHz. 
The polarization Stokes $Q$ and $U$ dust maps  are given by 

\begin{align}
	Q^{dust}(\nu,\hat{n}) =  f_{d}((\hat{n}))I^{dust}(\nu,(\hat{n}))\text{cos}(2\gamma (\hat{n})),
	\\
	U^{dust}(\nu,\hat{n}) =  f_{d}((\hat{n}))I^{dust}(\nu,(\hat{n}))\text{sin}(2\gamma (\hat{n})),
\end{align}
where $\gamma$ denotes the polarization angle and $f_{d} = \sqrt{Q^2+U^2}/I$ is the polarization fraction. 

We smooth all the $T,Q,U$ foreground maps with a Gaussian beam smoothing of $9^{\circ}$. The $\{Q,U\}$ maps are then converted over the full sky to get $E$ and $B$ foreground templates for all the 21 PRISM frequency channels at $\text{N}_{\text{side}} = 16$.

\subsection {Noise model}  Realistic experiments will always have detector noise contributions. To simulate the noise contribution, we generate random noise realizations for each CMB field for all 21 PRISM frequency bands. The noise sensitivity for temperature, denoted by $\sigma_{I}$ in the unit of $\mu$K.arcmin, is given in the third column of Table~\ref{tab:prism}. The polarization sensitivities  are obtained as $\sqrt{2}\sigma_{I}$.  The noise is assumed to be Gaussian, isotropic, and uncorrelated from pixel to pixel. An additional assumption is that there is no correlation between the Stokes $Q$ and $U$ maps. We generate random Gaussian realizations of $T,Q$ and $U$ maps using the rms values from Table~\ref{tab:prism}. The Stokes $Q$ and $U$ maps are then converted over the full sky to get the $E$ and $B$ mode maps.

\begin{table}
	\addtolength{\tabcolsep}{1pt}
	\renewcommand{\arraystretch}{1.2}
	\centering
	\begin{tabular}{ccc} 
		\hline
		\hline
		Freq.(GHz) & Beam FWHM  & CMB $\sigma_{I}$  \\
		& (arcmin) & ($\mu $K.arcmin)   \\
		\hline
		\hline
		21\hphantom{00} & \hphantom{0}15.36 & \hphantom{0}18.41  \\
		25\hphantom{00} & \hphantom{0}12.8 & \hphantom{0}12.88  \\
		30\hphantom{00} & \hphantom{0}11.32 & \hphantom{0}8.74  \\
		36\hphantom{00} & \hphantom{0}9.44 & \hphantom{0}6.13  \\
		43\hphantom{00} & \hphantom{0}8.88 & \hphantom{0}6.13  \\
		52\hphantom{00} & \hphantom{0}7.35 & \hphantom{0}4.29  \\
		62\hphantom{00} & \hphantom{0}5.12 & \hphantom{0}4.14  \\
		75\hphantom{00} & \hphantom{0}4.27 & \hphantom{0}3.22  \\
		90\hphantom{00} &  \hphantom{0}3.8 & \hphantom{0}2.14 \\
		108\hphantom{00} & \hphantom{0}3.16 & \hphantom{0}1.68  \\
		129\hphantom{00} & \hphantom{0}2.96 & \hphantom{0}1.68  \\
		155\hphantom{00} & \hphantom{0}2.48 & \hphantom{0}1.38  \\
		186\hphantom{00} & \hphantom{0}1.72 & \hphantom{0}3.06  \\
		223\hphantom{00} & \hphantom{0}1.44 & \hphantom{0}3.52  \\
		268\hphantom{00} & \hphantom{0}1.28 & \hphantom{0}2.3  \\
		321\hphantom{00} & \hphantom{0}1.04 & \hphantom{0}3.22  \\
		385\hphantom{00} & \hphantom{0}1.00 & \hphantom{0}3.52  \\
		462\hphantom{00} & \hphantom{0}0.84 & \hphantom{0}6.90  \\
		555\hphantom{00} & \hphantom{0}0.60 & \hphantom{0}35.29  \\
		666\hphantom{00} & \hphantom{0}0.52 & \hphantom{0}136.5  \\
		799\hphantom{00} & \hphantom{0} 0.44 & \hphantom{0} 807.1  \\ \hline
		\hline
	\end{tabular}
	\caption{PRISM beam and noise levels in units of $\mu$K.arcmin~\citep{2019arXiv190901591D}.}
	\label{tab:prism}
\end{table}

\section{Methodology} \label{Methods}
Combining the foreground templates with the noise and CMB maps described in Section~\ref{input maps}, we generate 21 input frequency maps. We reconstruct the posterior using these input maps in a model-independent manner. We remove the monopole and dipole contributions from the input maps before implementing our Gibbs-ILC algorithm, as they do not have any cosmological information. In our analysis, in order to sample the joint posterior density of CMB $P({\bf S}, C_{\ell}^X \vert {\bf D})$ our algorithm has 10 independent chains and each chain consists of 10000 Gibbs iterations. We initialized the Gibbs chains by randomly drawing the samples of $C_\ell^X$ from a uniform distribution of $\pm 3\Delta C_\ell^X $ which is consistent with Planck best-fit theoretical power spectrum, where  $\Delta C_\ell^X$ denotes the cosmic variance error.

\begin{figure*}
	\centering
	\includegraphics[scale=0.48]{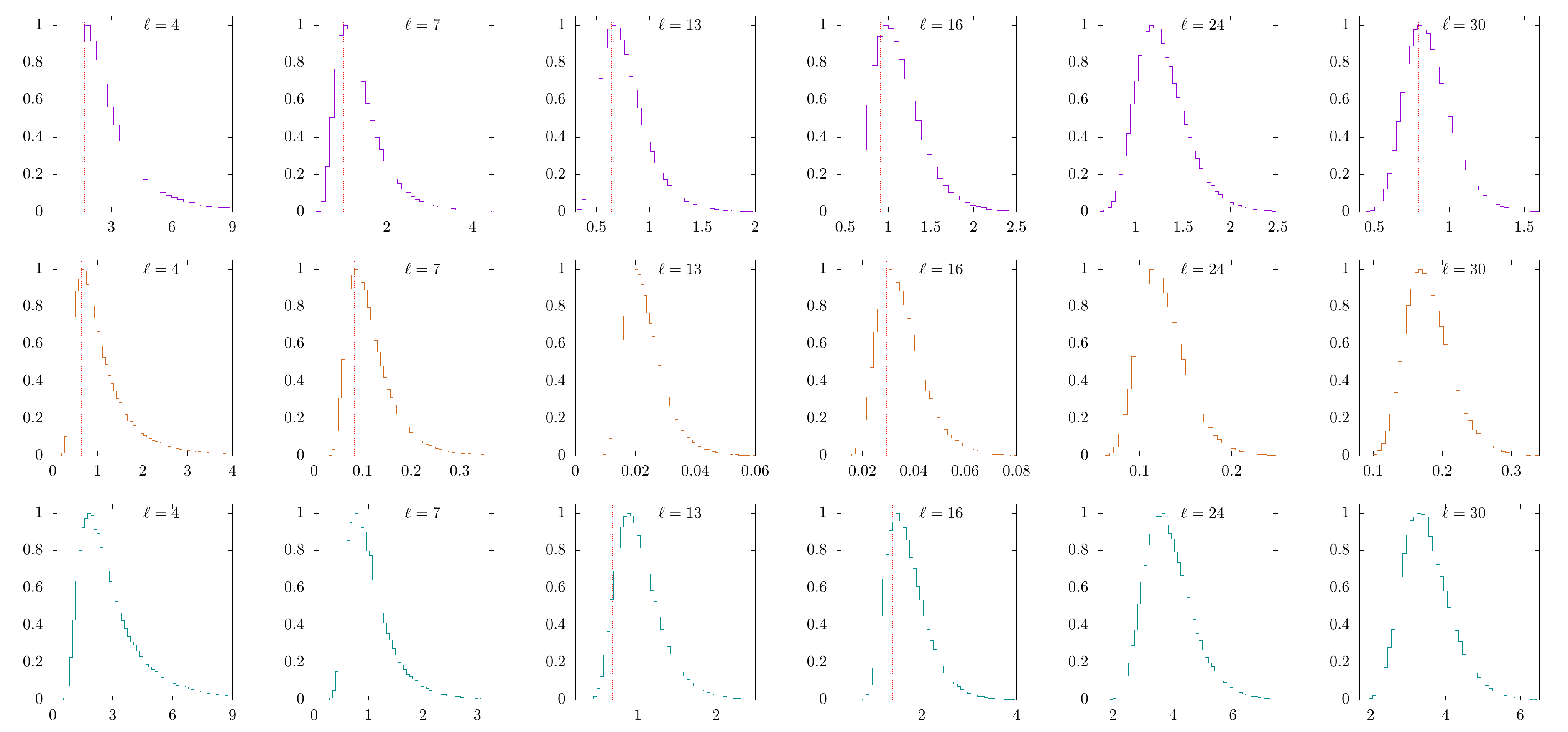}
	\caption{\small  We show some randomly chosen normalized histogram plots of the CMB cleaned angular power spectrum. In the top panel, we show histogram plots for CMB temperature, whereas the middle panel depicts the CMB $E$-mode . The bottom panel shows CMB $B$-mode in green. The horizontal axis for each subplots represents $\ell(\ell +1)C_\ell/2\pi$ in units of $10^3\mu K^2$, $10^{-1}\mu K^2$ and $10^{-4}\mu K^2$ for $T$, $E$ and $B$ cleaned angular power spectrum, respectively. The red vertical line represents the value of the fiducial angular power spectrum. The lower multipoles show asymmetry as they follow a inverse-gamma distribution. Since the degrees of freedom increase as we move towards higher multipoles, the density plots approach a symmetric distribution.\normalsize}
	\label{hist}
\end{figure*}

We sample a CMB theoretical $C_\ell$ and a cleaned CMB map following the procedure described in the section ~\ref{formalism} at each Gibbs step for every chain. For a given iteration, we use Eqn.~\ref{weight} to sample {\bf S}. The weights used in this equation are  obtained by utilizing Eqn.~\ref{weightnew} and the matrix elements of ${\bf A}$  are computed following the  Eqn.~\ref{hilc1} using the latest sampled $C_{\ell}^X$ and the corresponding sky $\hat C^X_\ell$. The Gibbs chain quickly stabilizes after an initial burn-in phase. In our analysis, we discard $100$ samples from the initial burn-in period. Therefore, we are left with net 99000 samples of $C_\ell^X$ and $\bf S$ for the analysis. Since we implement the ILC approach to mitigate the foregrounds, we would like to emphasise that the joint posterior density CMB signal over large scales reported in our analysis is insensitive to foreground modelling uncertainties.
\vspace{0cm}
\section{Results} \label{Sec:Results}

In this section, we discuss the results obtained after sampling the $T,E,B$ field by making use of Gibbs-ILC methodology and the corresponding theoretical angular power spectrum.

\subsection{Cleaned maps }\label{cm}

The marginalized probability densities of each of the 3072 reconstructed pixels are obtained by using the 99000 samples after the burn-in rejection of all the 10 Gibbs chains. The marginalized densities are then converted to normalized histograms by dividing with their modal values. By allocating the modal value of the corresponding histogram to each pixel value, we estimate the best-fit cleaned $T$ map. In the top panel of the first column of Fig.~\ref{cleanedmapsB}, we show the input CMB $T$ map for any random realization, and the cleaned best-fit CMB $T$ map for the same realization is shown in the second panel of the same column. It is evident from the figure that the best-fit map agrees very well with the input CMB map over the whole sky. The difference (residual map) between the best-fit cleaned $T$  map and the corresponding input map is shown in the middle panel (first column). The residuals are only of the order of $\lesssim 23.3\mu$K. For the quantitative study of the residual errors, we also compute the standard deviation map using all the 200 difference maps obtained from the 200 simulations. In the fourth panel (first column) of Fig.~\ref{cleanedmapsB}, we show the standard deviation map. In the central region of the galactic plane, one can see some pixels  ($\lesssim 15.8\mu$K) show only a minor level of residual contamination. Finally, the standard deviation weighted residual map in units of $\pm 3\sigma$ is shown in the last panel (first column). From our analysis we find that all the pixels in the standard deviation weighted residual map are within $\pm 3\sigma$. Thus, we can conclude that an efficient foreground removal and CMB $T$ signal reconstruction have been achieved.\\

The top panel of the second column of Fig.~\ref{cleanedmapsB} depicts the input CMB $E$ map, and the corresponding best-fit cleaned $E$ map is shown in the second panel of the same column. A visual inspection shows that an efficient foreground minimized map has been obtained. From the residual map shown in the middle panel (second column) it is evident that the residuals are only of the order $\lesssim 0.029\mu$K for CMB $E$ map. In the fourth panel (second column), we plot the standard deviation map obtained following a similar procedure as $T$ map. 
	We can see residual contamination of $\lesssim 0.028    \mu \textnormal{K}$ in the central and extreme left regions of the map. The residuals are along the galactic plane due to the presence of strong polarized foregrounds. From the standard deviation weighted residual map shown in the last panel (second column), we find that only 12 pixels are beyond the $\pm 3\sigma$ limits. Finally, in the third column of Fig.~\ref{cleanedmapsB}, we show the input, best-fit, residual,  standard deviation and standard deviation weighted residual $B$ maps from the first through fifth panels. From the morphological pattern and the map scale, we can infer that an efficient foreground reduced cleaned map has been estimated.   The upper and lower regions of the galactic plane show a minor level of contamination as evident from its standard deviation map. The reconstruction error due to residual foregrounds is not more than $0.0049 \mu \textnormal{K}$. Furthermore, from the standard deviation weighted residual map we also find that only 27 out of 3072 pixels exceed the $\pm 3\sigma$ bounds.

\subsection{Cleaned power spectrum }

\begin{figure*}
	\subfloat{\includegraphics[width=.5\linewidth]{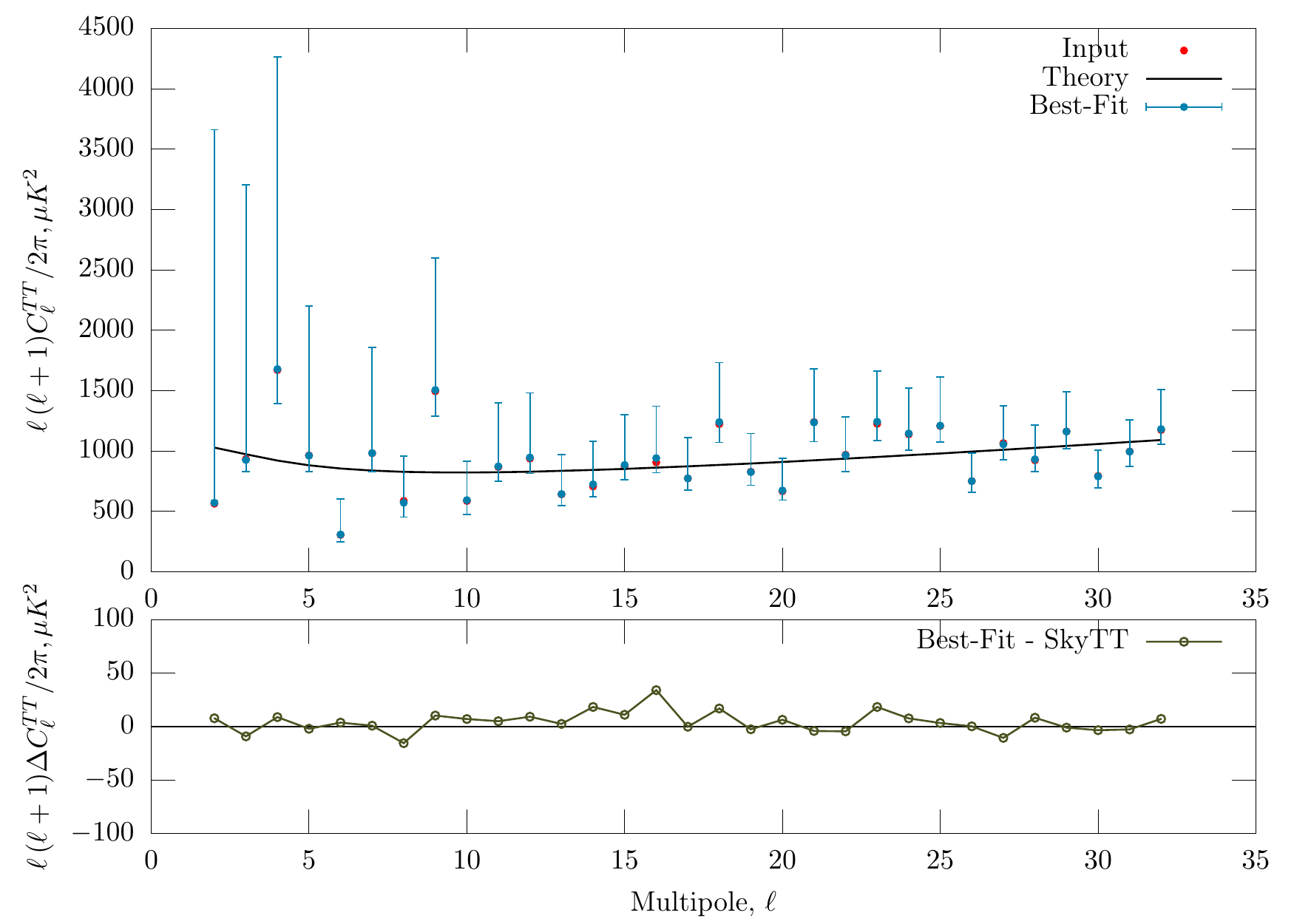}}\hfill
	\subfloat{\includegraphics[width=.5\linewidth]{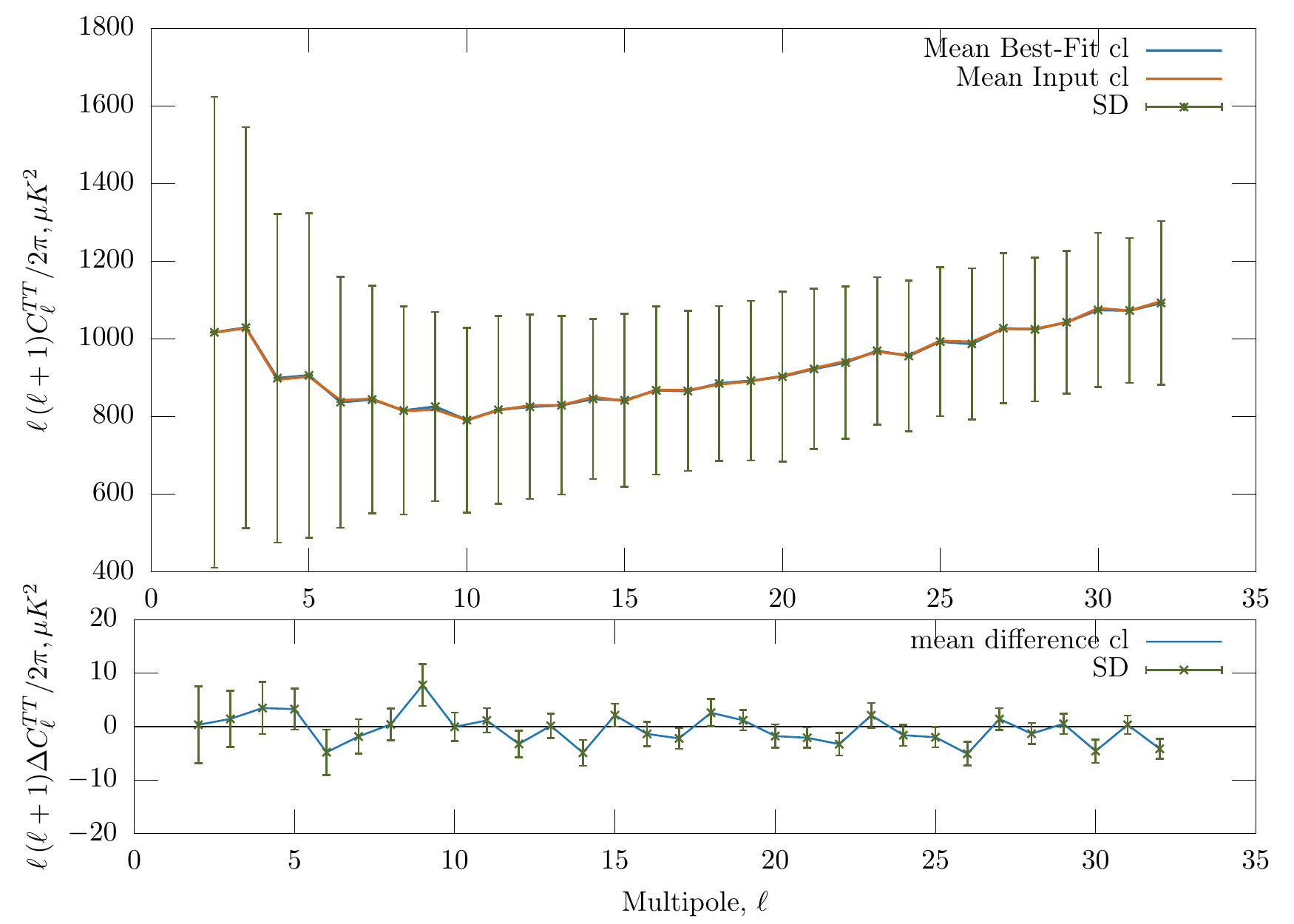}}\hfill       
	\caption{In the top left panel, we show the best-fit cleaned angular power spectrum in blue along with the asymmetric error bars obtained for an arbitrary sky realization using all the 99000 Gibbs samples. The input angular power spectrum is shown in red. In the bottom left panel, we show the difference between the best-fit angular power spectrum and  input sky $C^{TT}_{\ell}$. In the  top right panel, we show the mean cleaned angular power spectrum obtained from 200 simulations in blue along with the $1 \sigma$ standard error bars. The mean  input angular power spectrum is plotted in red. In the bottom right panel, we show the difference between the mean cleaned angular power spectrum and mean input $C^{TT}_{\ell}$ with $1 \sigma$ error bars. The error bars have been divided by a factor of $\sqrt{200}$.  From the figure, it is clear that an efficient foreground removal and  $C^{TT}_{\ell}$ reconstruction has been achieved.}\label{cleanedclT}
\end{figure*}

We estimate the marginalized posterior densities of reconstructed cleaned CMB $TT$ multipoles following a similar procedure outlined in section~\ref{cm}. The normalized histogram plots for some random multipoles are shown in top panel of Fig.~\ref{hist}. The bottom panel shows CMB $B$-mode in green. The horizontal axis for each subplots represents $\ell(\ell +1)C_\ell/2\pi$ in units of $10^3\mu K^2$, $10^{-1}\mu K^2$ and $10^{-4}\mu K^2$ for $T$, $E$ and $B$ cleaned angular power spectrum, respectively.   Using the modal values of the normalized histogram densities, we estimate the cleaned $TT$ power spectrum for all the multipoles ranging from $2-32$ along with the asymmetric error bars represented by blue. We show the reconstructed CMB theoretical $C^{TT}_{\ell}$ in the top left panel of Fig~\ref{cleanedclT} in the units of $\mu \text{K}^2$. The input $C^{TT}_{\ell}$ is plotted in red. The theoretical CMB temperature power spectrum is depicted as a black curve. The error bars show asymmetry on low multipoles, which, however, decreases as we go to higher multipoles. The cleaned $TT$ angular power spectrum matches well with the input power spectrum shown in the left top panel of Fig~\ref{cleanedclT}. The difference angular power spectrum $\Delta C^{TT}_{\ell}$ obtained from subtracting the best-fit angular power spectrum from the input sky $C^{TT}_{\ell}$ is shown in the left bottom panel. We show the mean best fit cleaned $TT$ angular power spectrum obtained from 200 simulations along with standard deviation error bars in the top right panel. In the bottom right panel, we show the mean difference angular power spectrum with the corresponding $1\sigma$ error bars. Here the error bars are divided by a factor of $\sqrt{200}$. From these difference plots we can infer that an efficient foreground removal has been achieved.  Thus from the analysis, we can conclude that our algorithm has accomplished efficient reconstruction across all the multipoles.

\begin{figure*}
	\includegraphics[width=.5\linewidth]{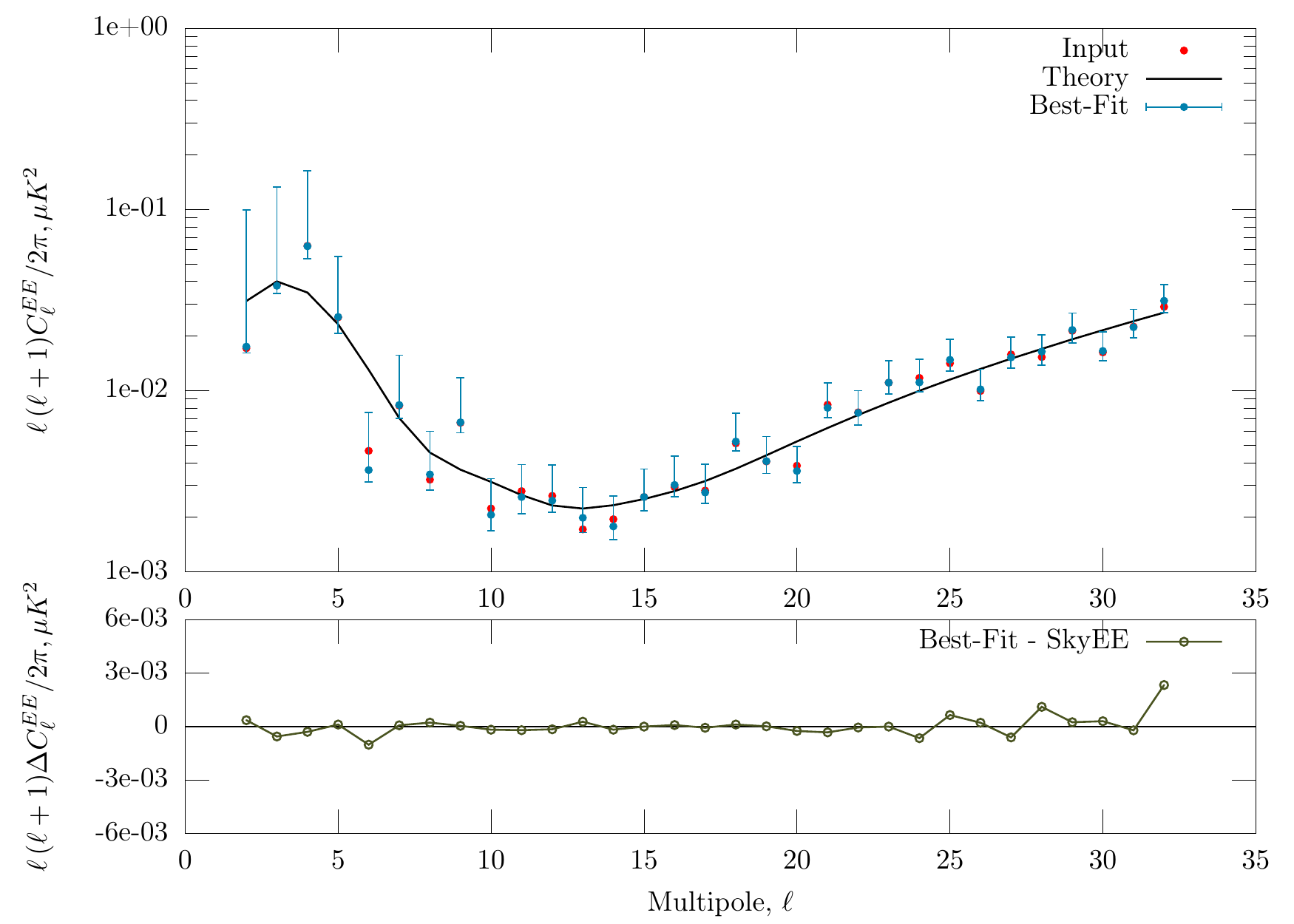}\hfill
	\includegraphics[width=.5\linewidth]{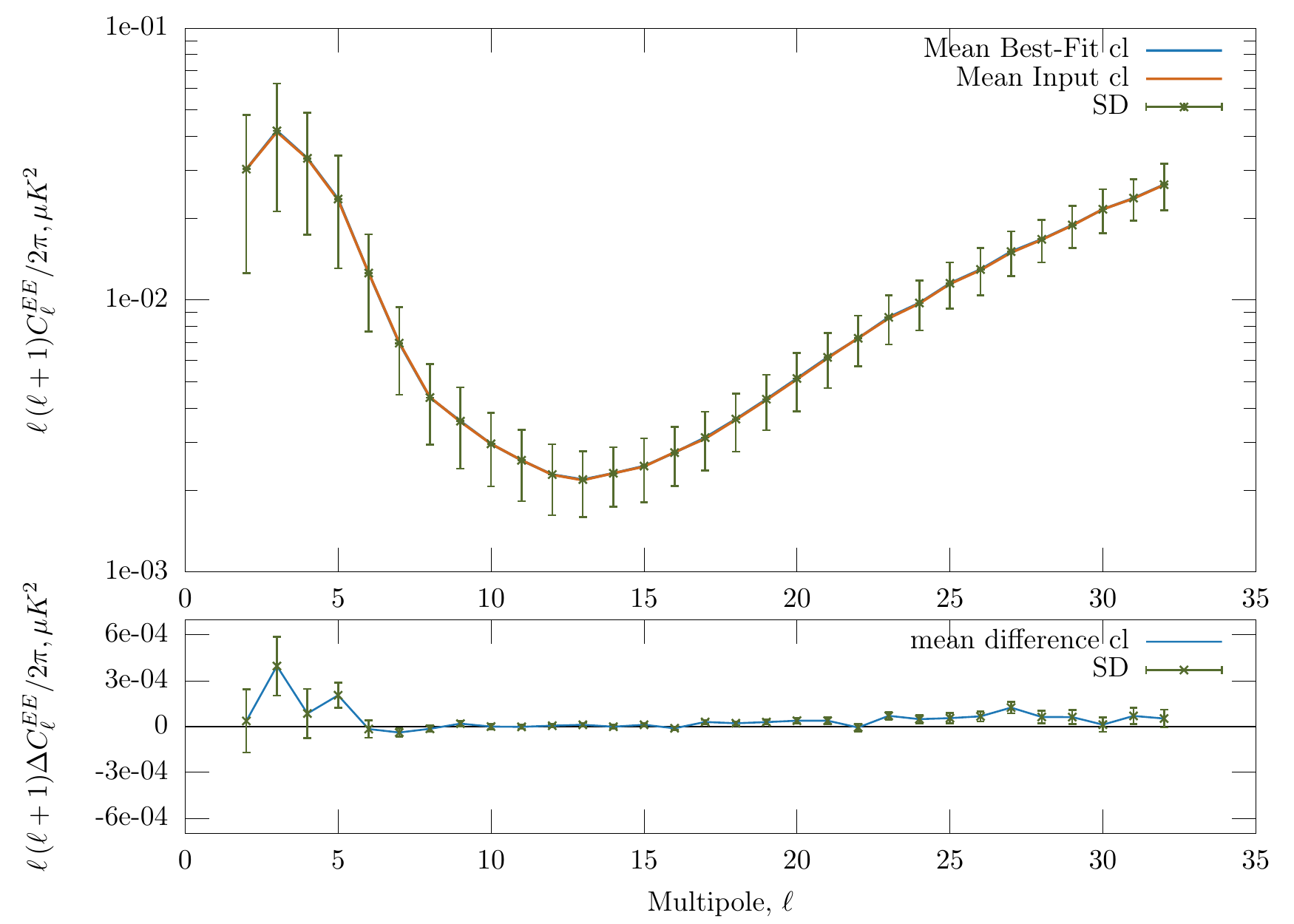}\hfill       
	\caption {Same as Fig.~\ref{cleanedclT} but for cleaned $C^E_{\ell}$. We see efficient foreground removal and reconstruction, except minor bias on multipoles ($\ell < 6$ ) as evident from the bottom right panel.} \label{cleanedclE}
\end{figure*}

The estimated cleaned $EE$ angular power spectrum from the histogram plots is shown in Fig.~\ref{cleanedclE}. The marginalised density plots are obtained similarly to the temperature case. In the top left panel of Fig.~\ref{cleanedclE}, we show the cleaned $EE$ angular power spectrum in blue points with asymmetric error bars for an arbitrary realisation. The input angular power spectrum corresponding to the same realisation is shown in red, whereas the theoretical CMB $EE$ angular power spectrum is plotted in black. We can see that the cleaned angular power spectrum overlaps with the input power spectrum for all the multipole ranges. The bottom left panel represents the difference between the cleaned angular power spectrum and the input $C^{EE}_{\ell}$. The difference plot shows a negligible residual bias in the power spectrum. Thus it is evident from Fig.~\ref{cleanedclE} that an efficient foreground removal has been achieved. We also estimate the mean cleaned $C^{EE}_{\ell}$ as shown in top right panel of Fig.~\ref{cleanedclE} with blue curve, obtained from 200 simulations along with the standard error bars. The mean input angular power spectrum is shown in red. The bottom panel shows the difference between the mean cleaned $C^{EE}_{\ell}$ and the input CMB $EE$ power spectrum. Clearly, we can infer that an efficient CMB $E$-mode reconstruction has been achieved.

\begin{figure*}
	\centering
	\subfloat{\includegraphics[width=.50\linewidth]{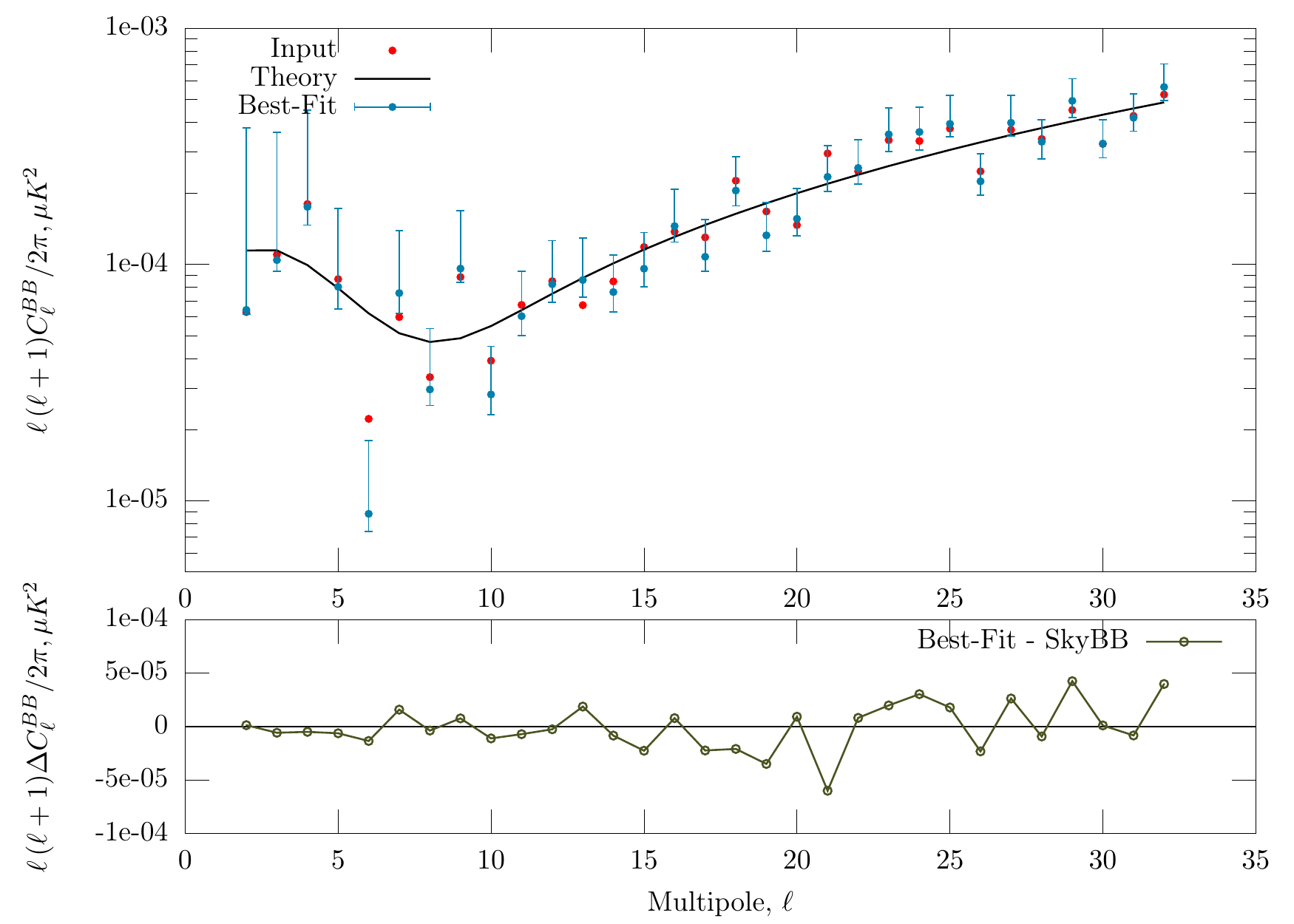}}\hfill
	\subfloat{\includegraphics[width=.50\linewidth]{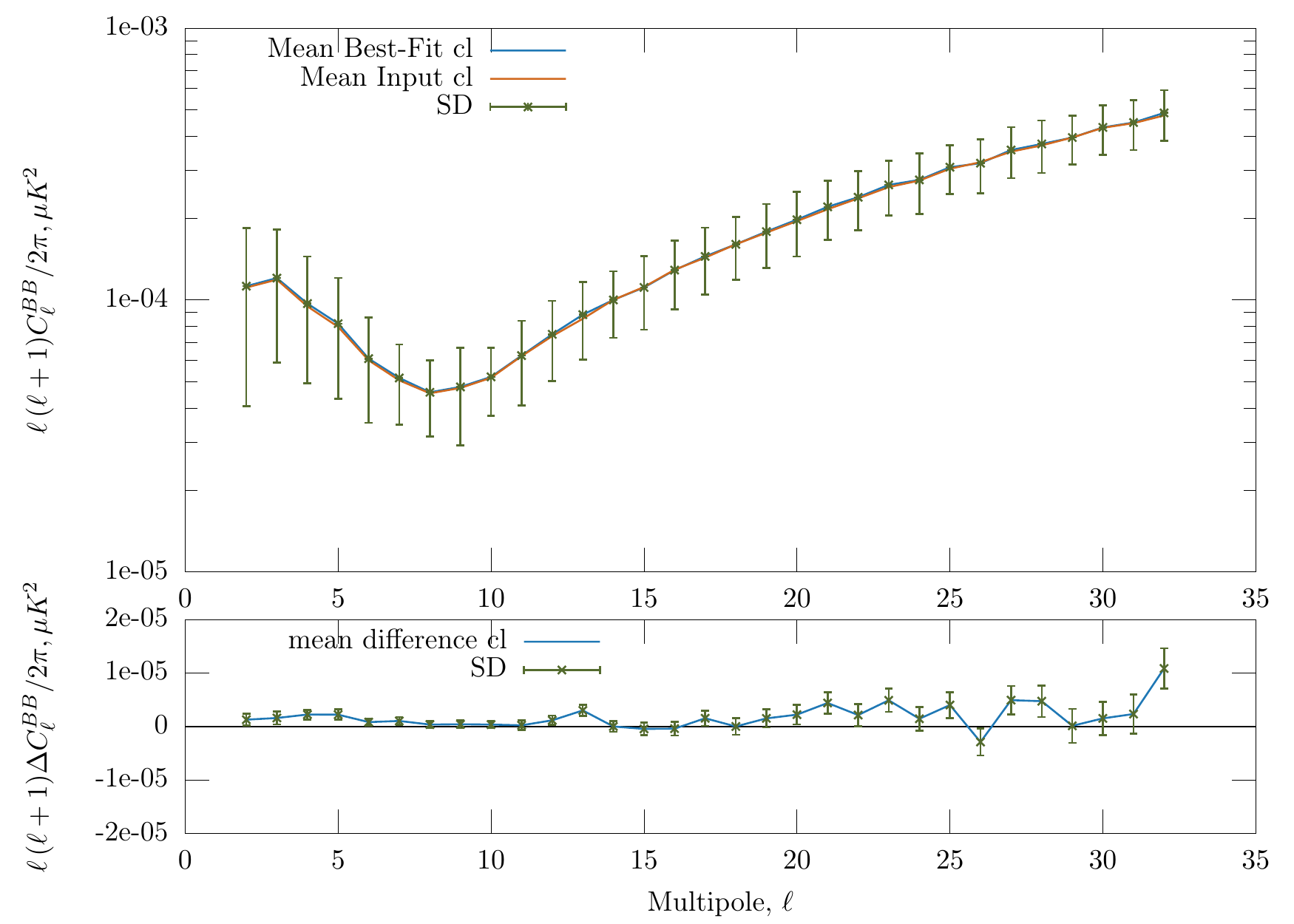}}\hfill
	\caption{Same as Fig.~\ref{cleanedclT} but for cleaned $C^B_{\ell}$. The cleaned angular power spectrum was estimated after subtracting the weighted auto noise power spectrum. \textcolor{black}{From the figure, it is evident that an efficient foreground removal has been accomplished.}}\label{cleanedclB}
\end{figure*}

In the bottom panel of Fig.~\ref{hist}, we show the marginalised density plots of $C^{BB}_{\ell}$ for some of the multipoles. The cleaned angular power spectrum is shown in the top left panel of Fig.~\ref{cleanedclB} in blue points, along with the asymmetric error bars. The input $C^{BB}_{\ell}$ is shown in red, whereas the theoretical $C^{BB}_{\ell}$ is plotted in black. The difference power spectrum $\Delta C^{BB}_{\ell}$ is represented in the bottom left panel. The top right panel of Fig.~\ref{cleanedclB} depicts the mean cleaned CMB $C^{BB}_{\ell}$  and the mean difference power spectrum along with the standard error bars is shown in the bottom right panel.  The error bars are of the order of $~10^{-5}$. From the figure it is evident that a good foreground removal and CMB $B$-mode reconstruction has been accomplished. \textcolor{black}{We would like to emphasise that our methodology can accurately reconstruct both
	the CMB $E$-mode ($\ell \approx 4$) and the CMB $B$-mode reionization bump ($\ell \simeq 4$)~\citep{2019BAAS...51g.286L} and the low power multipoles, which leads
	to a precise understanding of the physics of the reionization
	epoch. Moreover the efficient reconstruction of CMB $B$-mode
	angular power spectrum can also help us in extracting the
	signature of inflationary gravitational waves }.

From our analysis, we can conclude that our methodology can efficiently disentangle the realistic foregrounds and reconstruct the CMB temperature and the weak polarization signals. \textcolor{black}{ In  section~\ref{br}, we further utilize these reconstructed $TT,EE$ and $BB$ angular power spectra  for the estimation of cosmological parameters  which encodes valuable information about the  physics of early universe.}

\section{Blackwell Rao estimator and Parameter Estimation}
\label{br}

The ability to calculate the likelihood of any proposed CMB angular power spectrum given the data is essential for the accurate estimation of cosmological parameters. Though our methodology computes the posterior density of the theoretical CMB angular power spectrum, the underlying power spectrum is estimated discretely. A more precise assessment of the likelihood function of the CMB angular power spectrum can be made by applying the Blackwell-Rao theorem~\citep{PhysRevD.71.103002}. According to the theorem, an estimator can always be found
with a similar or higher efficiency than the initial estimator by using its conditional expectation with respect  to a sufficient statistic. The transformed estimator obtained by employing the Blackwell-Rao theorem is called the Blackwell-Rao estimator.

For the theoretical angular power spectrum $C_\ell^{X, th}$ corresponding to the CMB field $X$, the posterior distribution for the parameter set $\alpha$ can be computed with the Blackwell-Rao estimator using the Gibbs samples, $\hat C_\ell^{X(i)}$, of the reconstructed CMB power spectrum:
\begin{equation}
	P\left(\alpha\right) \approx {1\over N}\sum_{i=1}^{N} \mathcal{L}\left[{\widehat{C}}_\ell^{X\, (i)}|C_\ell^{th\, X}\left(\alpha\right)\right],   
\end{equation}\label{eq:br_posterior_tau}
\hspace{-0.12cm}where $N$ represents the total number of Gibbs samples obtained from all chains  after burn-in rejection and $\hat C_{\ell}^{X(i)}$ is the $i^{th}$ realization of the power spectrum obtained after excluding the burn-in samples from all Gibbs chains. The log-likelihood function is given by,
\begin{equation}
	\small
	-2\ln \mathcal{L}\left[{\widehat{C}}_\ell^{X(i)}|C_\ell^{X, th}\right]  = \sum_{\ell} (2\ell+1)\left[ \ln\left({C_\ell^{X, th}\over{\widehat{C}}_\ell^{X(i)}}\right) + {{\widehat{C}}_\ell^{X(i)}\over C_\ell^{X, th}} - 1\right].
	\small
\end{equation}

The fact that the estimated likelihood functions $\mathcal{L}$ are insensitive to the explicit foreground models is an intriguing feature of these functions. Therefore the likelihood functions are not susceptible to any modelling inaccuracies of the foregrounds. Furthermore, these likelihood functions are unaffected by the residual foregrounds. As our Gibbs-ILC approach uses a large number of input frequency bands, it leads to negligible foreground contamination in the cleaned CMB map and its theoretical angular power spectrum.

Since our analysis is on large angular scales, we focus on the primordial parameters, specifically the optical depth to reionization $\tau_{reio}$, the tensor to scalar ratio $r$ and the lensing amplitude $A_{lens}$. The posterior distribution for the parameter set $\alpha=\{\tau_{reio}, r, A_{lens}\}$ is obtained by sampling the parameter space with a Markov Chain Monte Carlo method (MCMC) method. We specifically use the modified version of Cobaya~\citep{2021JCAP...05..057T} to estimate the best-fit and $68\%$ limits of the cosmological parameters. Moreover, to generate theoretical CMB temperature and polarization power spectra, we use the latest version of the cosmological Boltzmann code CAMB~\citep{2000ApJ...538..473L}. We then sample the parameter space $\alpha$ by adopting flat priors. Moreover, we use the publicly accessible GetDist~\citep{2019arXiv191013970L} software package to statistically analyze the MCMC results. We sample the parameter space until the Gelman-Rubin convergence statistic~\citep{10.1214/ss/1177011136} satisfies $R - 1 < 0.01$.

\begin{figure}
	\includegraphics[scale=0.57]{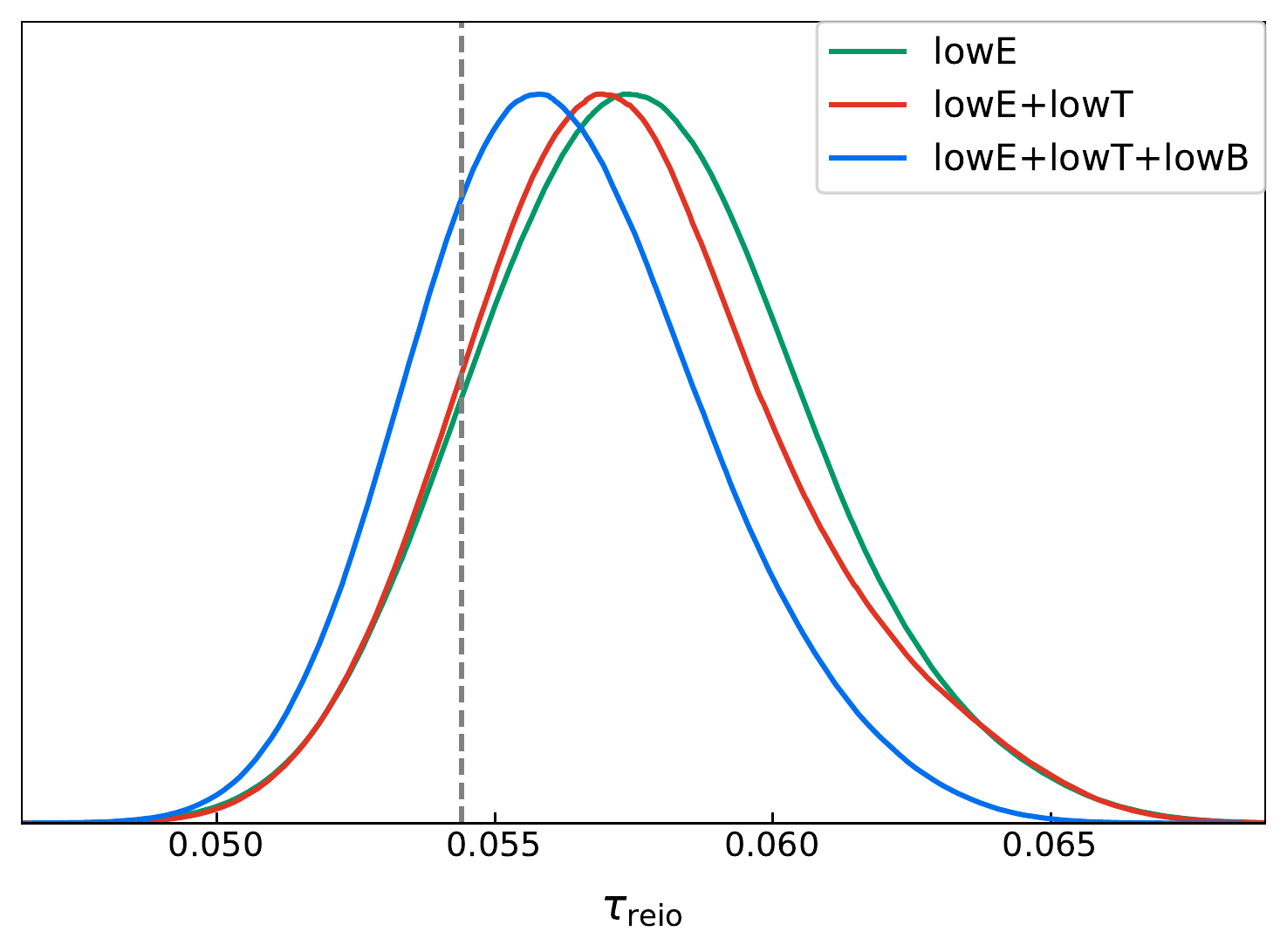}
	\caption{The figure shows the 1-dimensional marginalized distributions for the parameter optical depth to reionization, $\tau_{reio}$. The dashed line represents the fiducial value of $\tau_{reio}$. The constraints are tighter when the $EE$ power spectrum is integrated with $TT$ and $BB$ power spectrums.}
	\label{fig:tau_1d}
\end{figure}

\begin{figure}
	\includegraphics[scale=0.38]{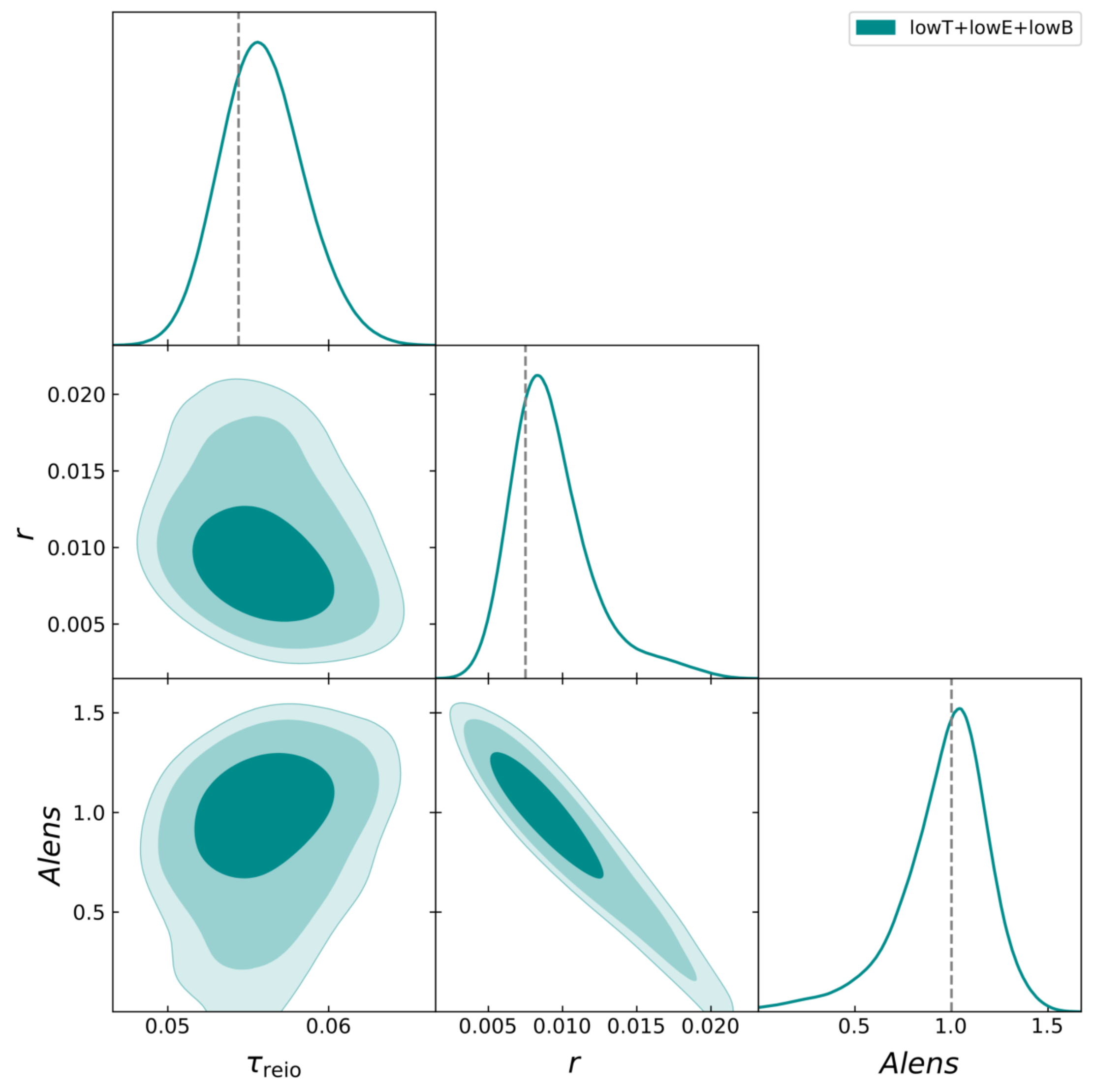}
	\caption{Figure showing 1-dimensional marginalized posterior distributions and 2-dimensional marginalized constraint contours for the parameters $\tau_{reio}$, $r$ and $A_{lens}$. The contours show $68\%$, $95\%$ and $99\%$ confidence regions. The dashed line represents the fiducial values of the parameters. We obtain the fiducial values of the parameters within $1\sigma$ standard deviation error. A noticeable correlation exists between the parameter pairs ($r$, $A_{lens}$). The quantitative results are summarized in Table~\ref{tab:best_fit_joint}.}
	\label{fig:joint_2d}
\end{figure}

The 1-dimensional marginalized distributions for the parameter, optical depth to reionization $\tau_{reio}$ is given in Fig.~\ref{fig:tau_1d}. To estimate the posterior distribution for $\tau_{reio}$, we fixed all cosmological parameters to the latest Planck 2018 results~\citep{2020A&A...641A...6P}, except for $\tau_{reio}$. From the quantitative results summarized in Table~\ref{tab:best_fit_tau}, it is interesting to note that there is a noticeable improvement in the parameter constraints when the CMB low-$\ell$ $EE$ power spectrum is integrated with CMB low-$\ell$ $TT$ and low-$\ell$ $BB$ power spectrum.
Apart from the information from $E$-mode, the CMB $BB$ power spectrum is also crucial for the estimation of $\tau_{reio}$ as it provides valuable insights on reionization physics from the reionization bump at multipole $\ell\sim4$~\citep{2019BAAS...51g.286L}. Thus, as seen from Table~\ref{tab:best_fit_tau}, with the addition of the CMB $BB$ power spectrum one can obtain tighter constraints on $\tau_{reio}$.
We also obtain the true value of $\tau_{reio}=0.0544$ within $1\sigma$ of the posterior distribution of $\tau_{reio}$. It is also noteworthy that there is a significant improvement in the constraints of $\tau_{reio}$ compared to its constraints from the latest Planck 2018 results ($\tau_{reio}=0.0544\pm0.0073$)~\citep{2020A&A...641A...6P}. The 1-dimensional marginalized posterior distributions and 2-dimensional marginalized constraint contours for the parameters using the joint analysis of the reconstructed power spectrum ($TT$, $EE$ and $BB$) are shown in Fig.~\ref{fig:joint_2d}. The contours show the $1\sigma$ region of $68\%$ confidence level, the $2\sigma$ region of $95\%$ confidence level, and $3\sigma$ region of $99\%$ confidence level, with the darker colour signifying the more probable results. Here we sample the parameters optical depth to reionization $\tau_{reio}$, the tensor to scalar ratio $r$ and the lensing amplitude $A_{lens}$ by adopting flat priors on them. The rest of the cosmological parameters are fixed to the latest Planck 2018 results~\citep{2020A&A...641A...6P}. The quantitative results from the MCMC analysis with $68\%$, $95\%$ and $99\%$ confidence levels are shown in Table~\ref{tab:best_fit_joint}.  From our analysis, we reconstructed the tensor-to-scalar ratio $r$ with a confidence level of $3\sigma$ and with a relative deviation of $0.3\sigma$ ( see Table~\ref{tab:best_fit_joint}). Our joint analysis for the parameter set $\alpha=\{\tau_{reio}, r, A_{lens}\}$ also obtains the true values within $1\sigma$ standard deviation error, establishing that we have performed an efficient foreground removal and CMB signal and angular power spectrum reconstruction. \textcolor{red}{}

\section{Discussion and Conclusions}
\label{disc}

The accurate measurement of the CMB $T,E$ and $B$ signals provide valuable information about the evolutionary history of our universe through the estimation of the cosmological parameters. The CMB temperature fluctuations have proven to be a crucial tool for probing the geometry~\citep{2020A&A...641A...6P} and formation of large-scale structures in the universe~\citep{2003AnPhy.303..203H}. Moreover, the physics of the ionized universe can throw light on various astrophysical processes like the formation of early stars. Detection of the yet unobserved $B$-mode signal will establish inflation as the basic mechanism for the origin of fluctuations and a nearly scale-invariant power spectrum. In this work,  we perform foreground removal and estimation of large angular scale full sky CMB $T,E$ and $B$ signals and their corresponding theoretical angular power spectrum. We used the proposed PRISM satellite mission to test and demonstrate our methodology. We incorporate the modified ILC method with the Gibbs sampling technique to draw samples from the joint density. 
An added advantage is that the large angular scale CMB-foreground chance correlations, which greatly affect the usual ILC are greatly reduced in the present work.
\begin{table}
	\addtolength{\tabcolsep}{13pt}
	\renewcommand{\arraystretch}{2}
	\centering
	\begin{tabular} { l  c}
		\hline
		\hline
		$\tau_{reio}$ &  Best-fit $\pm $68\% limits\\
		\hline
		\hline
		{lowE} & $0.0571^{+0.0031}_{-0.0027}         $\\
		
		{lowE+lowT} & $0.0568^{+0.0029}_{-0.0026} $\\
		
		{lowE+lowT+lowB} & $0.0558^{+0.0026}_{-0.0023}     $\\
		\hline
		\hline
	\end{tabular}
	\caption{ The best-fit values with $68\%$ intervals for the parameter $\tau_{reio}$. }
	\label{tab:best_fit_tau}
\end{table}

\begin{table}
	\addtolength{\tabcolsep}{4pt}
	\renewcommand{\arraystretch}{1.8}
	\centering
	\begin{tabular} { l c c}
		\hline
		\hline
		Parameter & Fiducial Value  & lowT+lowE+lowB  \\
		\hline
		\hline
		{\boldmath$\tau_\mathrm{reio}$} & $0.0544$&  $0.0555^{+0.0024+0.0053+0.0074}_{-0.0028-0.0048-0.0063}        $\\
		
		{\boldmath$r       $} & $0.0075$  & $0.0081^{+0.0028+0.0075+0.0111}_{-0.002-0.0039-0.0044}$\\
		
		{\boldmath$A_{lens}          $} & $1.0$  &  $1.051^{+0.25+0.44+0.50}_{-0.16-0.52-0.85}      $\\
		\hline
		\hline
	\end{tabular}
	\caption{ Best-fit values with 68\%, 95\% and 99\% confidence level constraints on the parameters. }
	\label{tab:best_fit_joint}
\end{table}

From our analysis, we infer that the reconstructed mean cleaned angular power spectra and the mean input CMB power spectra agree well with each other.
Since the polarization signal is weak, it is susceptible to significant detector noise contamination resulting in the foreground and noise bias in the cleaned angular power spectrum as well as foreground residuals in the cleaned maps. In order to remove the detector noise bias from the cleaned $B$-mode power spectrum, we perform the noise bias correction by subtracting the weighted noise auto power spectrum from the cleaned power spectrum. We find that the resultant cleaned angular power spectrum matches well with the theoretical angular power spectrum for all multipole ranges. From our analysis, we can conclude that our methodology can efficiently disentangle the realistic foregrounds and reconstruct the CMB temperature and the weak polarization signals. \textcolor{black}{We would like to emphasize that our methodology can accurately reconstruct both the CMB $E$-mode ($\ell \approx 4$) and the CMB $B$-mode reionization bump ($\ell \simeq 4$) and the low power multipoles, which leads to a precise understanding of the physics of the reionization epoch. Moreover, the efficient reconstruction of CMB $B$-mode angular power spectrum can also help us in extracting the signature of inflationary gravitational waves}.

After obtaining the reconstructed cleaned CMB angular power spectrum, we also estimate the parameters and their corresponding error limits using the MCMC sampling method. Apart from the information from $E$-mode, the CMB $BB$ power spectrum is also vital for the estimation of $\tau_{reio}$ as it provides valuable insights on reionization physics from the reionization bump at multipole $\ell\sim4$~\citep{2019BAAS...51g.286L}. Thus, with the integration of the CMB $BB$ power spectrum to $EE$ and $TT$ power spectrum, as expected, we do see the constraints on $\tau_{reio}$ are tighter than the latest Planck 2018 results. Our joint analysis using reconstructed cleaned CMB angular power spectrum ($TT, EE$ and $BB$), for the parameters $\tau_{reio}, r$ and $A_{lens}$, also obtains the true values within $1\sigma$ standard deviation error, establishing that we have performed an efficient foreground removal and CMB signal and angular power spectrum reconstruction. From our analysis, we also reconstructed the tensor-to-scalar ratio $r$ with a confidence level of $3\sigma$ and with a relative deviation of $0.3\sigma$. From the accurate reconstruction of CMB angular power spectra and precise estimation of cosmological parameters, we can infer that our methodology is reliable and can efficiently mitigate the foregrounds.  

In the current work, we demonstrate an analysis pipeline which inputs foreground contaminated CMB maps at several observed frequencies and estimates reliable CMB signal, its angular power spectra along with their likelihood functions. The method then estimates the relevant cosmological parameters using the marginalized likelihood functions of CMB spectra. An important aspect of the CMB component analysis is that one uses a CMB signal reconstruction methodology that can provide error estimates on the estimated signal and angular power spectrum. For our analysis, we employ a unique Gibbs-ILC method, which has twofold advantages. First, the methodology is independent of the foreground model, and second, in this context, we can also estimate the joint posterior distributions of the cleaned signal and theoretical angular power spectrum. Since the Gibbs-ILC method provides the marginalized probability density functions of the signal and angular power spectrum, it is natural to utilize these products for cosmological parameter estimation. Our method is unique in the sense that it is foreground model-independent in nature on one side and Bayesian on the other. The Bayesian nature makes the method useful for cosmological parameter estimation. The Bayesian properties also imply that foreground reconstruction errors are nicely propagated in the final cosmological parameter estimation. We would also like to mention that our foreground removal and parameter estimation are modular in nature so that, if required, they can be modified as need be raised.
It also has the added advantage that one can implement the same pipeline when the real data becomes available. In a future article, to extract and estimate
all fundamental cosmological parameters we will extend
our methodology on high-resolution CMB data.

\section*{ACKNOWLEDGEMENTS}
We use the publicly available HEALPix package~\citep{Gorski_2005} for the analysis of this work (http://healpix.sourceforge.net).

\section{Data Availability}
The data underlying this article will be shared on reasonable request
to the corresponding author.
\bibliographystyle{mnras}
\bibliography{ms}

\label{lastpage}
\end{document}